\documentclass[aps,prl,twocolumn,preprintnumbers,amsmath,amssymb,superscriptaddress]{revtex4-2}
\usepackage{graphicx}
\usepackage[caption=false]{subfig}
\usepackage{epsfig}
\usepackage{dcolumn}
\usepackage{bm}
\usepackage{color}
\usepackage{float}
\usepackage[colorlinks]{hyperref}
\usepackage{verbatim}
\usepackage{algorithm}
\usepackage{algorithmic}
\usepackage{tikz}
\usetikzlibrary{quantikz}
\usepackage{xr}
\include{ref}
\usepackage{titletoc}

\hypersetup{citecolor = blue}

\def\be{\begin{equation}}       \def\ee{\end{equation}}
\def\bea{\begin{eqnarray}}      \def\eea{\end{eqnarray}}
\def\ba{\begin{array}}
\def\ea{\end{array}}
\def\bnum{\begin{enumerate} }
\def\enum{\end{enumerate}}

\def\=>{\Rightarrow}
\def\>{\rightarrow}

\def\eye2{Fathbb{I}}

\newcommand{\no}{\nonumber}

\renewcommand{\>}{\rangle}

\renewcommand{\rm}[1]{\mathrm{#1}}

\usepackage{listings}
\usepackage{color}
\definecolor{lightgray}{gray}{1}

\begin{document}

\title{Discrete time crystal enabled by Stark many-body localization}

\author{Shuo Liu}
\affiliation{Institute for Advanced Study, Tsinghua University, Beijing 100084, China}
\affiliation{Tencent Quantum Laboratory, Tencent, Shenzhen, Guangdong 518057, China}
\author{Shi-Xin Zhang}
\email{shixinzhang@tencent.com}
\affiliation{Tencent Quantum Laboratory, Tencent, Shenzhen, Guangdong 518057, China}
\author{Chang-Yu Hsieh}
\affiliation{Tencent Quantum Laboratory, Tencent, Shenzhen, Guangdong 518057, China}
\author{Shengyu Zhang}
\affiliation{Tencent Quantum Laboratory, Tencent, Shenzhen, Guangdong 518057, China}
\author{Hong Yao}
\email{yaohong@tsinghua.edu.cn}
\affiliation{Institute for Advanced Study, Tsinghua University, Beijing 100084, China}

\begin{abstract}
Discrete time crystal (DTC) has recently attracted increasing attention, but most DTC models and their properties are only revealed after disorder average. In this Letter, we propose a simple disorder-free periodically driven model that exhibits nontrivial DTC order stabilized by Stark many-body localization (MBL). We demonstrate the existence of DTC phase by analytical analysis from perturbation theory and convincing numerical evidence from observable dynamics. The new DTC model paves a new promising way for further experiments and deepens our understanding of DTC. Since the DTC order doesn't require special quantum state preparation and the strong disorder average, it can be naturally realized on the noisy intermediate-scale quantum (NISQ) hardware with much fewer resources and repetitions. Moreover, besides the robust subharmonic response, there are other novel robust beating oscillations in Stark-MBL DTC phase which are absent in random or quasi-periodic MBL DTC.
\end{abstract}

\date{\today}
\maketitle

{\bf Introduction:}
Spontaneous symmetry breaking (SSB) is one of the most important concepts in modern physics. Various phases of matter and phase transitions can be described by SSB mechanism; for example, the formation of crystals is the result of spontaneously breaking continuous spatial translational symmetry. Inspired by this notion, Wilczek proposed the intriguing concept of ``time crystal'' which spontaneously breaks continuous time translational symmetry \cite{PhysRevLett.109.160401, PhysRevLett.109.163001, PhysRevLett.111.250402} and various no-go theorems \cite{PhysRevLett.111.070402, nozieres_time_2013, PhysRevLett.114.251603} have established since then that the continuous time crystal would not exist. However, Floquet systems, quantum systems subject to periodic driving, can exhibit discrete time translational symmetry breaking (DTTSB) \cite{PhysRevLett.116.250401, PhysRevLett.117.090402, PhysRevB.93.245146, PhysRevB.94.085112} and have attracted considerable research interest \cite{PhysRevA.91.033617, PhysRevLett.117.090402, PhysRevLett.116.250401, PhysRevLett.118.030401, PhysRevB.95.214307, PhysRevLett.120.040404, PhysRevLett.120.110603, zhu_dicke_2019, PhysRevLett.123.210602, PhysRevLett.125.060601, yao_classical_2020, PhysRevResearch.3.L042023}. The given observable in the DTC phase can develop persistent oscillations whose period is an integer multiple of the driving period. Recently, DTC has been experimentally realized in programmable quantum devices with periodic driving \cite{choi_observation_2017, zhang_observation_2017, mi_time-crystalline_2021, doi:10.1126/science.abk0603, zhang_digital_2022, doi:10.1126/sciadv.abm7652}.

Due to the existence of periodic driving, energy is no longer conserved in a Floquet system. Thus, in the absence of any other local conservation laws, a generic system will absorb energy from the periodic driving, ultimately heating to infinite temperature. The thermalization of many-body Floquet systems implies that any local physical observable becomes featureless at late times \cite{PhysRevX.4.041048, PhysRevE.90.012110, ponte_periodically_2015}. Therefore, strong disorder is required \cite{PhysRevLett.116.250401, PhysRevLett.117.090402, PhysRevB.93.245146, PhysRevB.94.085112, PhysRevLett.114.140401, ponte_periodically_2015, PhysRevLett.115.030402, abanin_theory_2016} to realize MBL that exhibits emergent local integrals of motion \cite{PhysRevLett.111.127201, PhysRevB.90.174202} and prevents absorption of heat from periodic driving. However, to investigate the DTC behavior, we have to average the observable dynamics over a great number of different disorder instances, requiring more quantum resources and severely restricting the efficient experimental study of DTC.

Besides DTC stabilized by MBL phase, so-called prethermal DTC phase exists without the need of MBL. Under some conditions, the dynamics of the many-body Floquet system can be thought of as being generated by an effective time-independent ``prethermal Hamiltonian'', $H_{\rm{eff}}$. The Floquet system can then display DTC dynamics upon starting from certain low-temperature symmetry-breaking initial states of $H_{\rm{eff}}$ within an exponential heating time window \cite{abanin_exponentially_2015, mori_rigorous_2016, kuwahara_floquet-magnus_2016, abanin_effective_2017, abanin_rigorous_2017}, realizing prethermal DTC \cite{PhysRevX.7.011026, PhysRevA.91.033617, PhysRevB.96.094202, PhysRevX.10.021046, doi:10.1126/science.abg8102, PhysRevB.103.224311, PhysRevA.99.033618}.

Recently, in kicked PXP model \cite{doi:10.1126/science.abg2530, maskara_discrete_2021}, the discrete time crystal order enabled by quantum many-body scars \cite{turner_weak_2018, PhysRevLett.122.040603, PhysRevLett.123.147201, serbyn_quantum_2021} with N$\acute{\rm{e}}$el state as the initial state had been identified, which is strongly reminiscent of prethermal DTC. (See \cite{collura_discrete_2022} for another similar mechanism enabling sub-Hilbert space DTC behavior in which the DTC lifetime can be enhanced with dynamical freezing \cite{PhysRevB.97.245122, PhysRevX.11.021008, Haldar_2022}) The fidelity
$
F_{n} = \vert \langle Z_{2} \vert U_{F}^{n} \vert Z_{2} \rangle \vert^{2}
$
is used to characterize the dynamics in this model,
where $\vert Z_{2} \rangle$ is the initial N$\acute{\rm{e}}$el state and $U_{F}$ is the Floquet evolution unitary. When $n$ is even, $F_{n} > 0$ and when $n$ is odd, $F_{n}=0$, this corresponds to the subharmonic response with a timescale $T_{s}=2$. After a long enough time, the fidelity will decay to zero finally and stay featureless, and this corresponds to the prethermal timescale $T_{p}$. Between the two timescales, there are another two novel timescales that are not reported in DTC or prethermal DTC phases before. One is the emergent beating timescale $T_{b}$, the fidelity at even periods exhibits a beating oscillation which comes from the overlap between the N$\acute{\rm{e}}$el initial state and the lowest lying excited states of $H_{\rm{eff}}$. The beating timescale $T_{b} \propto \Delta^{-1}$ where $\Delta$ is the gap in the Floquet spectrum. The other is the timescale $T_{g}$ that is set by the inverse energy splitting in the ground state manifold and $T_{g} \propto e^{N}$ where $N$ is the system size. After driving cycles in the order of $T_{g}$, the fidelity at even periods $F_{2n}$ decreases, while simultaneously the fidelity at odd periods $F_{2n+1} $ increases. However, different from prethermal DTC phases,  these phenomena strongly depend on special N$\acute{\rm{e}}$el initial states where highly accurate quantum state preparation is required.

An extremely important and exciting direction is to identify a clean Floquet system, i.e. without strong disorder, that exhibits a nontrivial DTC phase with no dependence on initial states. To stabilize this intrinsically dynamical phase, MBL is extremely important as discussed above. It has long been established that the quantum systems may enter MBL phases in the presence of sufficiently strong random disorder \cite{PhysRevLett.95.206603, basko_metalinsulator_2006, PhysRevB.75.155111, PhysRevB.75.155111, PhysRevB.77.064426, PhysRevB.81.134202, cuevas_level_2012, PhysRevLett.109.017202, PhysRevLett.110.067204, PhysRevLett.110.260601, PhysRevLett.111.127201, PhysRevB.90.174202}, quasi-periodic potential \cite{PhysRevB.87.134202, PhysRevLett.115.230401, PhysRevLett.121.206601, PhysRevLett.122.170403, zhang_strong_2019, PhysRevB.102.224203} or linear Zeeman field \cite{PhysRevLett.122.040606, PhysRevB.103.L100202, van_nieuwenburg_bloch_2019, khemani_localization_2020, Bhakuni_2020} in one-dimensional (1D) systems. The third one is called Stark MBL \cite{PhysRevLett.122.040606, PhysRevB.103.L100202, van_nieuwenburg_bloch_2019, khemani_localization_2020, Bhakuni_2020, sarkar_protecting_2022}. By intuition, we may construct a clean many-body Floquet system utilizing Stark MBL to stabilize DTC order.

In this Letter, we propose a clean kicked Floquet model inspired by Stark MBL. It has various nontrivial and interesting properties, including robust subharmonic response as conventional DTC and other novel timescales similar to those reported in kicked PXP model and Rydberg atom experiments \cite{doi:10.1126/science.abg2530, maskara_discrete_2021}. The subharmonic response is robust against imperfection and doesn't depend on special initial states, which is a signal for nontrivial DTC phase. Since the existence of DTC order doesn't depend on the strong disorder, it can be naturally realized on the quantum hardware relying on fewer quantum computational resources and experimental trials.

{\bf Model:}
The Floquet unitary for the model considered in the Letter reads:
\begin{eqnarray}
U_{F}=U_{2}U_{1}=e^{-iH_{2}}e^{-i(\frac{\pi}{2}-\epsilon)H_{1}},
\end{eqnarray}
where
\begin{eqnarray}
H_{1}&=&\sum_{j}X_{j}, \\
H_{2}&=&J_{z}\sum_{j}(j+1)Z_{j} Z_{j+1} + W \sum_{j}j Z_{j}.
\label{zzH2}
\end{eqnarray}
$X$ and $Z$ are Pauli matrices, $\epsilon$ is the imperfection in the driving. $H_{1}$ is the kicked term, when $\epsilon = 0$, $U_{1} = \prod_{j}X_{j}$ and all the spins flip exactly; $H_{2}$ has two terms: one is a linear Zeeman field for Stark MBL; the other one is a linear $zz$ interaction. The linear term for interaction is important to stabilize DTC similar to the case discussed in \cite{mi_time-crystalline_2021} (see also in the Supplemental Materials (SM) \footnote{See Supplemental Material at URL (to be added) for details, including the following: (1) the perturbation theory for dominant-subleading $\pi$-pair pattern, (2) analysis of different timescales, (3) the Fourier transform on the fidelity dynamics, (4) the numerical results for different imperfection $\epsilon$, (5) level statistics for many-body localization, (6) the numerical results with different choices of $J_z$ and $W$, (7) the dynamics of autocorrelator, (8) phase transition and the impact of the generic interaction, (9) the realization of DTC on quantum computers, and Refs. \cite{santagati_witnessing_2018, liu_probing_2021, PhysRevLett.110.084101, PhysRevLett.100.070502, jian_long-range_2015}.}). The reason is that the linear Zeeman field, as well as the Stark MBL, will be suppressed by the imperfection $\epsilon$. Therefore we need some nonuniform interaction to stabilize the MBL phase. Different from the random MBL DTC case where a strong disorder in $zz$ interaction is added, we here instead also use a linear $zz$ interaction.

When $\epsilon=0$, the quasi-eigenstates of $U^{0}_{F}$ can be written as:
\begin{eqnarray}
\vert \pm \rangle &=& \frac{1}{\sqrt{2}}(e^{-i\frac{H_{2}(z)}{2}} \vert z \rangle \pm e^{-i\frac{H_{2}(-z)}{2}} \vert -z \rangle ), \label{superposition}
\end{eqnarray}
whose eigenvalues are,
\begin{eqnarray}
U^{0}_{F} \vert \pm \rangle &=& \pm e^{-i \frac{H_{2}(z)+H_{2}(-z)}{2}} \vert \pm \rangle
\end{eqnarray}
respectively, where $\vert z \rangle $ is the product state and $\vert -z \rangle = \prod_{j}X_{j} \vert z \rangle$. $\vert + \rangle$ and $\vert - \rangle$ form a so-called $\pi$-pair in which quasi-eigenenergy difference equals $\pi$. For simplicity, we use quasi-eigenenergy $\varepsilon_{F}$ of $\vert + \rangle$ to represent this $\pi$-pair. And for any product state $\vert z \rangle$, it can be represented by a superposition of a $\pi$-pair of quasi-eigenstates,
\begin{eqnarray}
\vert \pm z \rangle = \frac{1}{\sqrt{2}} e^{i \frac{H_{2}(\pm z)}{2}} (\vert + \rangle \pm \vert - \rangle).
\end{eqnarray}
Therefore, there is a trivial subharmonic response  in $\epsilon=0$ limit,
\begin{eqnarray}
U_{F}^{2} \vert z \rangle = U_{F} \vert -z \rangle = \vert z \rangle,
\end{eqnarray}
where we have ignored the global phase. Accordingly, the local physical observables, such as $\langle Z(t) \rangle$, develop persistent oscillations whose periods are twice as the driving period, and the discrete time translational symmetry spontaneously breaks. However, this DTC order depends on fine-tuning of parameters $\epsilon=0$. To establish a nontrivial DTC phase, the subharmonic response must be robust against imperfection $\epsilon$. When $\epsilon \neq 0$, the quasi-eigenstates can not be analytically exactly tracked, we use perturbation theory and numerical results below to show that the subharmonic response is robust against imperfection $\epsilon$.

{\bf Observable:} To describe the dynamics of the kicked model and diagnose the DTC phase, we need to utilize suitable observable. Due to the linear Zeeman field and linear $zz$ interaction, different spatial sites are not equivalent anymore. Therefore, we don't choose the non-equal time spin-spin correlation on site $N/2$ commonly used in previous DTC works \cite{PhysRevLett.118.030401} and instead use more representative site-averaged observables. Spurred by \cite{maskara_discrete_2021}, we define two types of fidelity, one is the fidelity for a given initial state:
\begin{eqnarray}
F_{n} = \vert \langle \psi_{i} \vert U_{F}^{n} \vert \psi_{i} \rangle \vert^{2},
\end{eqnarray}
where $\vert \psi_{i} \rangle$ is the initial state. The other is the state-averaged fidelity
\begin{eqnarray}
\bar{F}_{n} = \frac{1}{2^N} \sum_{\{z\}}\vert \langle z \vert U_{F}^{n} \vert z \rangle \vert ^{2},
\end{eqnarray}
where the sum is over all possible $2^N$ product states $\vert z \rangle$. We can also utilize site-averaged spin autocorrelator as the observable $	\overline{\langle Z(0)Z(n) \rangle} = \frac{1}{N}\sum_{j} \langle Z_j(0)Z_j(nT) \rangle$,  and the dynamical behaviors are qualitatively the same as the fidelity (see the SM for details \cite{Note1}).


{\bf Analysis of different timescales:}
We observe three different timescales for our Floquet model which are similar to those reported in kicked PXP model. Because the third timescale $T_{g} \propto e^{N}$ is exponential with the system size, we show the dynamics of a small system (3 sites) firstly to demonstrate all the three timescales in the dynamics. The results are summarized in Fig. \ref{fig:L3TsTbTg}.

\begin{figure}[t]\centering
	\includegraphics[width=0.45\textwidth]{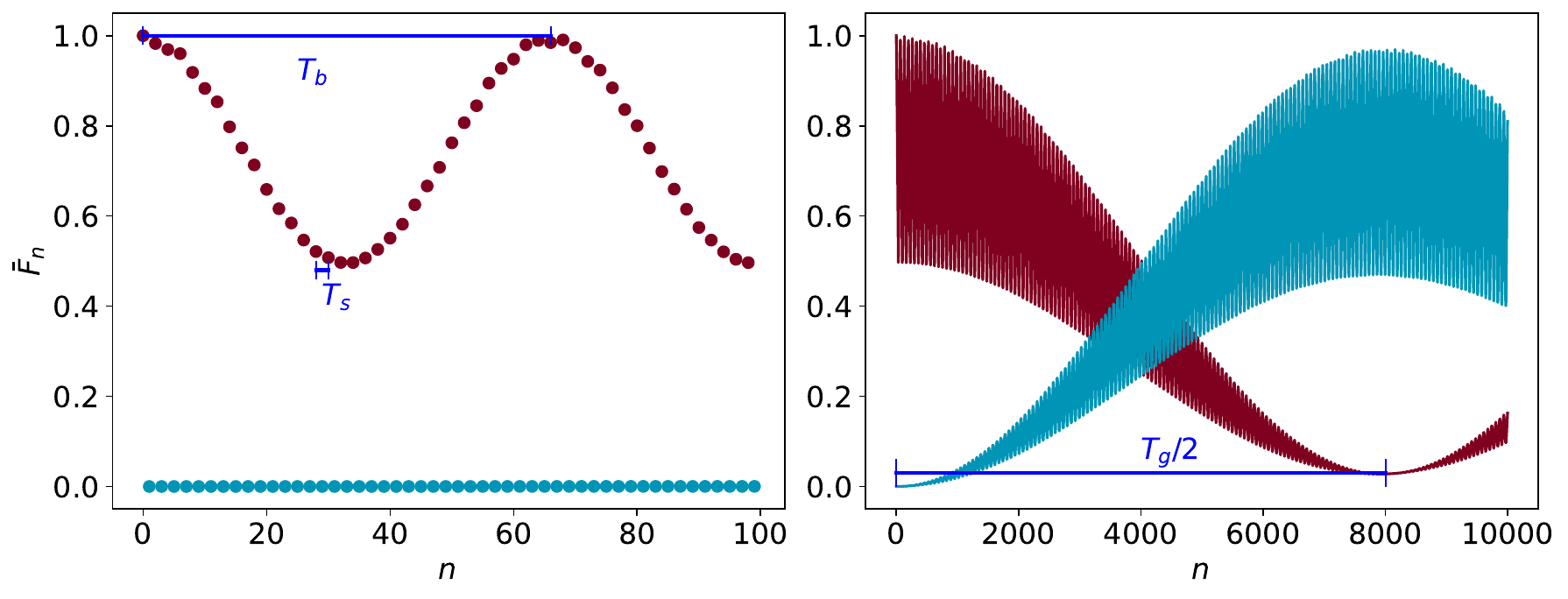}
	\caption{Three timescales for the dynamics of the state-averaged fidelity $\bar{F}_{n}$: when $n$ is even, $\bar{F}_{n}>0$ and when $n$ is odd, $\bar{F}_{n}=0$, which correspond to the subharmonic response $T_{s}=2$; besides this subharmonic response, there are a beating timescale $T_{b}\approx 66$ and a third timescale $T_{g} \approx 16000$ ($N=3; J_{z}=\frac{\pi}{2N}; W=5.0; \epsilon=0.05$).}
	\label{fig:L3TsTbTg}
\end{figure}

 To understand these three timescales, we consider a product state $\vert z \rangle$ and the corresponding state-dependent fidelity firstly. When $\epsilon \neq 0$, although the product state $\vert z \rangle$ can not be represented by a superposition of one $\pi$-pair of quasi-eigenstates, we can still do decomposition in the eigenspace of $U_{F}$. Based on the perturbation theory, the quasi-eigenstate of $U_{F}$ ($\epsilon \neq 0$) can be written as a superposition of the quasi-eigenstates of $U_{F}^{0}$ ($\epsilon=0$) with the similar quasi-eigenenergies. For a given product state $\vert z \rangle$, the corresponding original $\pi$-pair of quasi-eigenstates and the most related quasi-eigenstates to the first-order perturbation form a Hilbert subspace, and $\vert z \rangle$ roughly live in this subspace.

We utilize perturbation theory to locate the subspace where the product state $\vert z \rangle$ lives. We use $\vert \pm \rangle$ to represent the original $\pi$-pair related to $\vert z \rangle$ when $\epsilon=0$ and the quasi-eigenenergy is $\varepsilon_{F}$, i.e. $U_{F}^{0} \vert \pm \rangle = \pm e^{-i\varepsilon_{F}} \vert \pm \rangle$. By intuition, the dimension of the subspace is determined by the number of quasi-eigenstates of $U^{0}_{F}$ which have the similar quasi-eigenenergies with $\vert \pm \rangle$. In the SM \cite{Note1}, we show that if there is only one $\pi$-pair of quasi-eigenstates $\vert \pm^{\prime} \rangle$ with $\varepsilon_{F}^{\prime} = \varepsilon_{F} \,\, \rm{or} \,\, \varepsilon_{F} + \pi \,\, (\rm{mod} \, 2\pi)$, $\vert \pm \rangle$ and $\vert \pm^{\prime} \rangle$ form a subspace with dimension of 4 and $\vert z \rangle$ roughly lives in this subspace. Equivalently, if we check the overlaps between $\vert z \rangle$ and quasi-eigenstates of $U_{F}$, there is an obvious dominant-subleading $\pi$-pair pattern, see Fig. \ref{fig:overlap}(a). Even if there is no exactly matching quasi-eigenstates of $U_{F}^{0}$, as long as there is one special $\pi$-pair of quasi-eigenstates of $U_{F}^{0}$ which has the closer quasi-eigenenergy with $\vert \pm \rangle$ than all other eigenstates, the dominant-subleading $\pi$-pair pattern for the decomposition of  $\vert z \rangle$ still exists. On the contrary, if there are several $\pi$-pairs of $U_{F}^{0}$ have the same closest quasi-eigenenergy difference with $\vert \pm \rangle$ as $\delta \varepsilon_{F} = \vert \varepsilon_{F}-\varepsilon_{F}^{\prime} \vert = \delta $, the dominant-subleading $\pi$-pair pattern vanishes and we can only see one $\pi$-pair (the original one) with dominant overlap, see Fig. \ref{fig:overlap}(b).

\begin{figure}[t]\centering
	\includegraphics[width=0.45\textwidth]{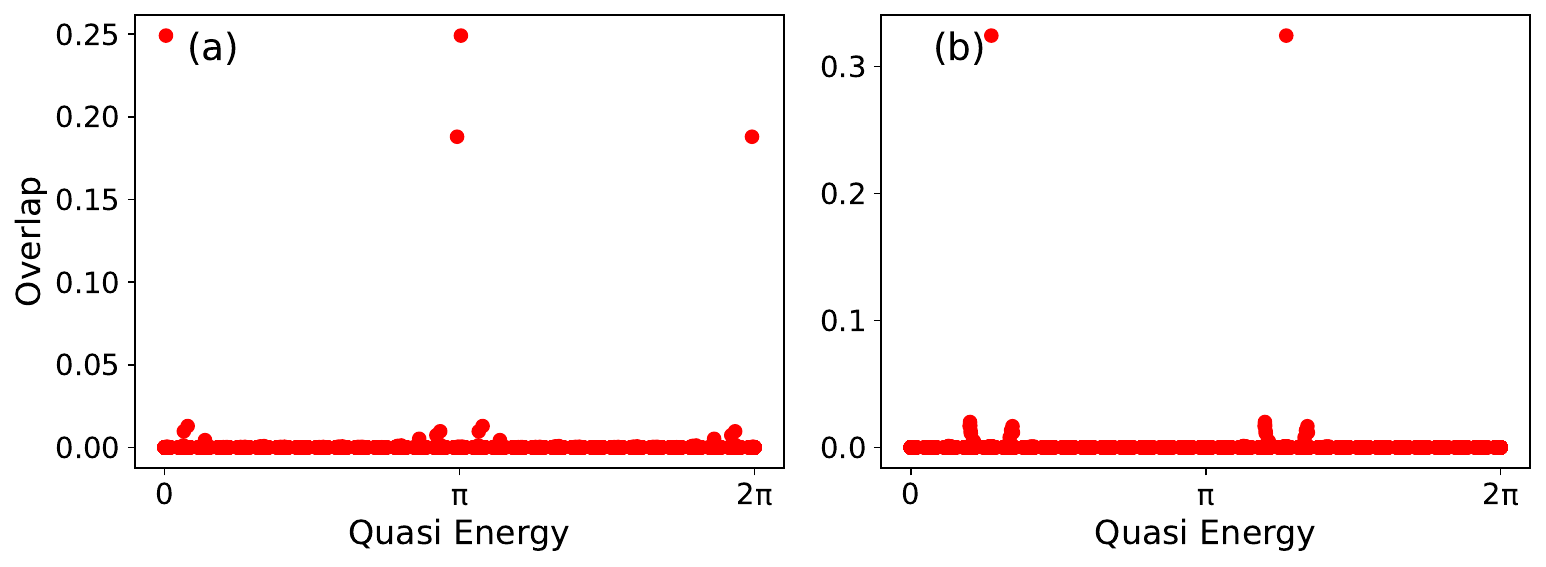}
	\caption{Overlaps with quasi-eigenstates of $U_{F}$: $N=15; J_{z}=\frac{\pi}{2N}; W=5.0; \epsilon=0.05$. $(a)$, the product state $\vert 000000000000000 \rangle$ roughly lives in a subspace with dimension of 4 and we can see the obvious dominant-subleading $\pi$-pair pattern; $(b)$, there is only one obvious $\pi$-pair in the decomposition of the product state $\vert 001100110011001 \rangle$.}
	\label{fig:overlap}
\end{figure}

When $\epsilon=0$, the product state $\vert z \rangle$ can be represented by an equal weight superposition of a $\pi$-pair of quasi-eigenstates of $U_{F}^{0}$ and this induces the subharmonic response as discussed above. When $\epsilon$ is small, as long as the product state $\vert z \rangle$ can still be represented by a superposition of several $\pi$-pairs of quasi-eigenstates of $U_{F}$ and the weights of the two quasi-eigenstates in any $\pi$-pair have the same absolute value, as guaranteed by the perturbation theory, the $T_s=2$ subharmonic response still exists.

The beating timescale $T_{b}$ is caused by the quasi-eigenenergy difference between different quasi-eigenstates (see the SM for details \cite{Note1}). Consider a three-site subsystem of a product state $\vert z \rangle$, $\vert S_{j-1}, S_{j}, S_{j+1} \rangle$, $\vert z \rangle$ and $\vert -z \rangle \, (\vert -S_{j-1}, -S_{j}, -S_{j+1} \rangle )$ can be combined into a $\pi$-pair with quasi-eigenenergy equals $\frac{H_{2}(z)+H_{2}(-z)}{2}$ when $\epsilon =0$. If we flip spin $S_{j}$ of $\vert z \rangle$ and $\vert -z \rangle$, there are two new product states $\vert z^{\prime} \rangle \, (\vert S_{j-1}, -S_{j}, S_{j+1} \rangle)$ and $\vert -z^{\prime} \rangle \, (\vert -S_{j-1}, S_{j}, -S_{j+1} \rangle)$. And they form a new $\pi$-pair $\vert \pm^{\prime} \rangle$ with quasi-eigenenergy $\varepsilon_{F}^{\prime} $. The quasi-eigenenergy difference between the two $\pi$-pairs is
\begin{eqnarray}
\delta \varepsilon_{F} = 2J_{z}\; \left \vert jS_{j-1}+(j+1)S_{j+1} \right\vert \,\, (\rm{mod} \,\, 2\pi).
\end{eqnarray}
When $S_{j-1}=-S_{j+1}$, the quasi-eigenenergy difference after flipping the spin $S_{j}$ equals $2J_{z}$; when $S_{j-1}=S_{j+1}$, the quasi-eigenenergy difference after flipping the spin $S_{j}$ equals $2(2j+1)J_{z}$. Suppose $J_{z}=\frac{\pi}{2N}$ and assume $N$ be odd for simplicity, considering the spin configuration around a fixed site $\frac{N-1}{2}$ (the middle site), the product states can be divided into two parts: when $S_{\frac{N-1}{2}-1}=S_{\frac{N-1}{2}+1}$, as discussed above, there are two $\pi$-pairs related by a Pauli-X matrix at site $\frac{N-1}{2}$ and the quasi-eigenenergy difference is equal to $\pi$. And the quasi-eigenstates decomposition of the product state includes a dominant $\pi$-pair and a subleading $\pi$-pair, see Fig. \ref{fig:overlap}(a).  We call these product states ``good initial states". The dynamics of these good states have a dominant beating timescale $T_{b}$ determined by the quasi-eigenenergy difference between dominant and subleading $\pi$-pairs and $T_{b}$ fits well with the perturbative predictions (see the SM for details \cite{Note1}), see Fig. \ref{fig:state0}. When $S_{\frac{N-1}{2}-1}=-S_{\frac{N-1}{2}+1}$, there is no $\pi$-pair $\vert \pm^{\prime} \rangle$ with quasi-eigenenergy difference $\delta \varepsilon_{F} =0  \,\, \rm{or} \,\, \pi$ with $\vert \pm \rangle$. As shown in Fig. \ref{fig:overlap}(b), there is no dominant-subleading $\pi$-pair pattern. We call these product states ``bad initial states''. Although there is no obvious subleading $\pi$-pair, there are still many $\pi$-pairs with small overlaps and different quasi-eigenenergy differences with $\vert \pm \rangle$, thus we can see many Fourier peaks in Fig. \ref{fig:state6-and-average}(b)  (see the SM for details \cite{Note1}).


\begin{figure}[t]\centering
	\includegraphics[width=0.45\textwidth]{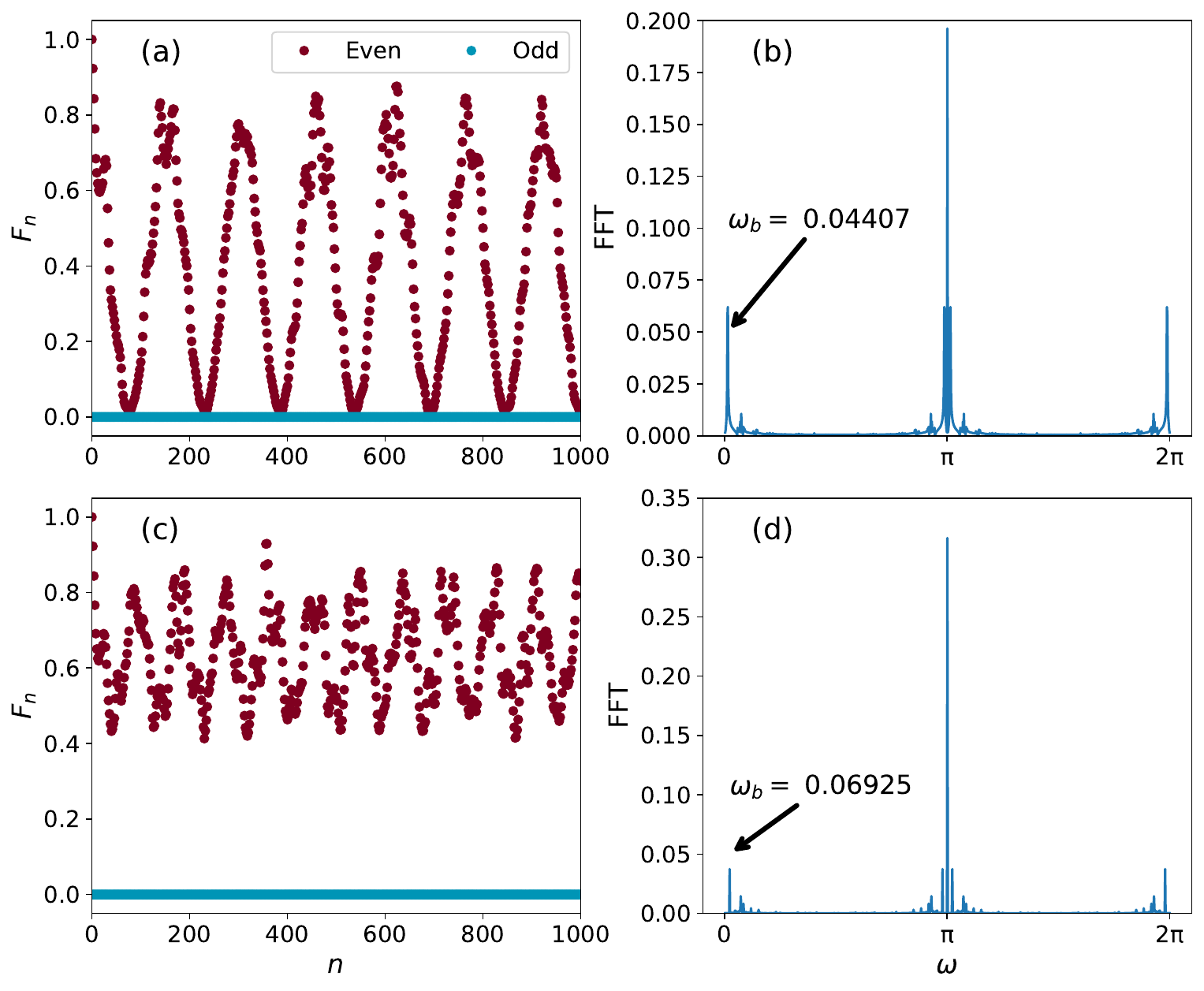}
	\caption{Dynamics of the product state $\vert 000000000000000 \rangle$ which is a ``good initial state'' ($N=15; W=5.0; \epsilon=0.05$). In (a)(b), $J_{z}=\frac{\pi}{2N}$ and in (c)(d), $J_{z}=\frac{\pi-0.05}{2N}$. In both cases, there is a dominant beating timescale.}
	\label{fig:state0}
\end{figure}

\begin{figure}[t]\centering
	\includegraphics[width=0.45\textwidth]{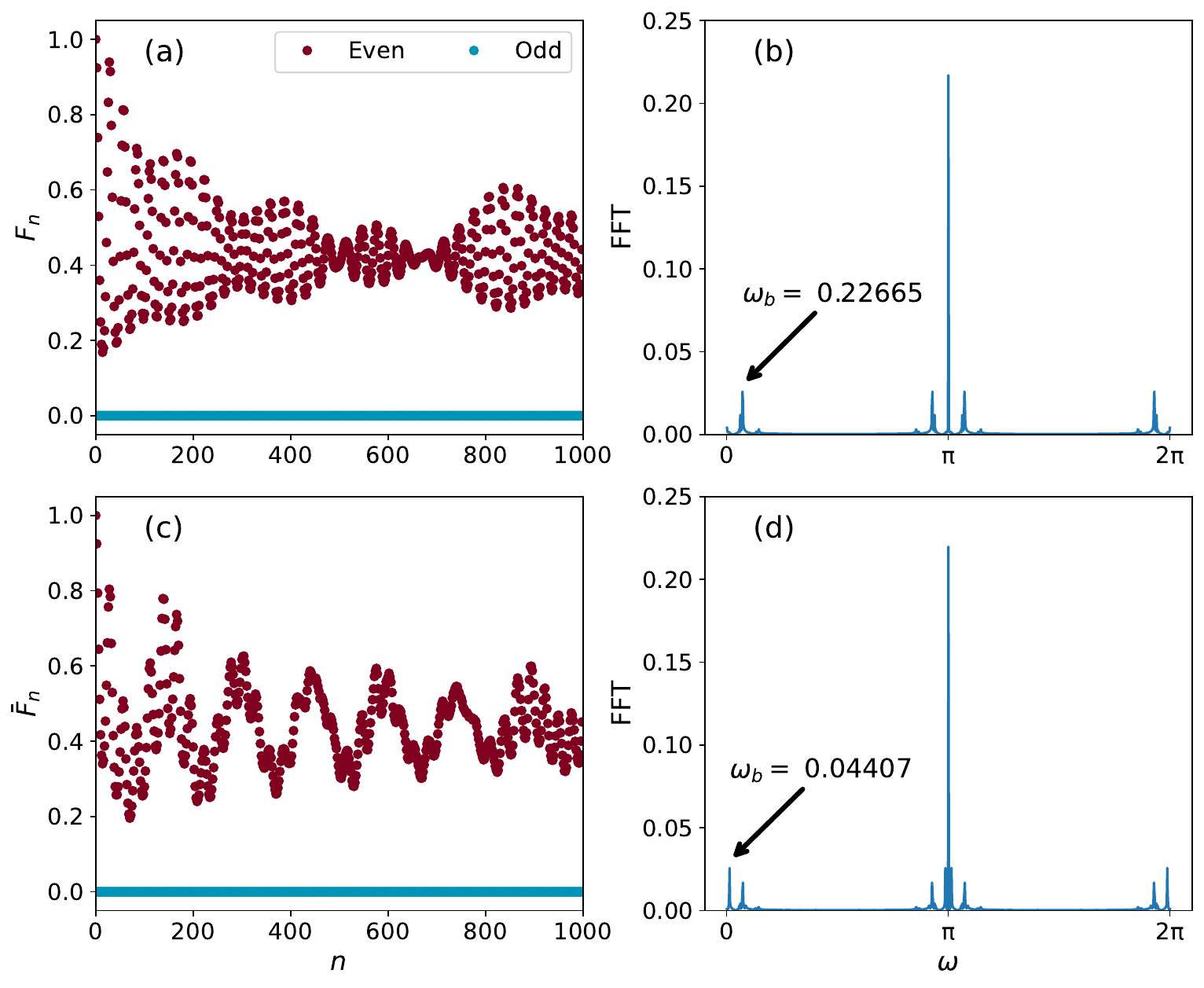}
	\caption{(a)(b): dynamics of the product state $\vert 001100110011001 \rangle$ ($N=15; J_{z}=\frac{\pi}{2N}; W=5.0; \epsilon=0.05$). Such a ``bad initial state" gives another beating timescale. (c)(d): dynamics of the state-averaged fidelity $\bar{F}_{n}$ ($N=15; J_{z}=\frac{\pi}{2N}; W=5.0; \epsilon=0.05$). The dominant beating timescale $T_{b}=\frac{2\pi}{\omega_{b}}$  is determined by the ``good initial states''.}
	\label{fig:state6-and-average}
\end{figure}

Now we investigate the robustness of beating timescale by considering non-perfect  $J_{z}$, for example, $J_{z}=\frac{\pi-0.05}{2N}$. For a $\pi$-pair of quasi-eigenstates of $U_{F}^{0}$ combined by ``good initial states'', although there is no other $\pi$-pairs of quasi-eigenstates of $U_{F}^{0}$ with $\delta \varepsilon_{F}=0 \,\, \rm{or} \,\, \pi$, as long as only one $\pi$-pair of quasi-eigenstates has closer quasi-eigenenergy with $\vert \pm \rangle$ than all other quasi-eigenstates, the dominant-subleading $\pi$-pair pattern still exists and there is a dominant beating timescale, see Fig. \ref{fig:state0}(d).

We can use a more general state-averaged observable, the state-averaged fidelity, to describe the dynamics of the many-body Floquet system. The quasi-eigenenergy corrections due to the first-order perturbation to all ``good initial states'' are the same, in the case considered here, $T_{b} \approx \frac{\pi}{\epsilon \sin(\frac{N-1}{2}W)}$. Although there is another beating timescale for ``bad initial states'', the dominant beating timescale for state-averaged quantities is determined by ``good initial states'', see Fig. \ref{fig:state6-and-average}(d) (see the SM for analytical analysis \cite{Note1}). Additionally, there is no scaling relation between $T_{b}$ and the system size $N$. Note that the classification of ``good initial state'' and ``bad initial state'' is only for the beating timescale $T_{b}$, all initial states show a robust subharmonic response $T_{s}=2$ breaking discrete time translational symmetry.

In terms of the timescale $T_{g}$, it is induced by the tiny quasi-energy splitting of a given $\pi$-pair, i.e. the quasi-eigenenergy difference between $\vert + \rangle$ and $\vert - \rangle$ equals $\pi+\delta$. This quasi-energy mismatch $\delta$ due to finite size effect induces the third timescale as $T_{g} \propto \frac{1}{\delta}$  and $\delta \propto e^{-N}$ (see the SM for details \cite{Note1}). As discussed in \cite{khemani_brief_2019}, the existence of the third timescale $T_{g}$ depends on the order of the two limits: (a), $\rm{lim}_{t\rightarrow \infty} \rm{lim}_{N\rightarrow \infty}$ and (b), $\rm{lim}_{N\rightarrow \infty} \rm{lim}_{t\rightarrow \infty}$. (a) characterizes the “intrinsic” quench dynamics of this phase. In (a), we will never reach times of $O(e^{N})$ and the third oscillation timescale $T_{g}$ ($\propto e^{N}$) vanishes. And we can only observe the subharmonic response and beating oscillation out to $t \rightarrow \infty$.

$J_z$ with dependence on the system size $N$ investigated in this Letter is designed to facilitate the analytical understanding of the beating timescale and is not necessary for the realization of a stable DTC phase. More numerical results with size independent $J_{z}$ can be found below and in the SM \cite{Note1}.

{\bf Phase transition:} In this section, we investigate the DTC order and choose the general size independent $J_z$.
To stabilize DTC phase, MBL is extremely important as discussed above. As shown in  Fig. \ref{fig:his-r-N13-main}, the distribution of level spacing ratio \cite{PhysRevB.75.155111} gradually crosses from the Poisson limit to the Gaussian orthogonal ensemble (GOE) with increasing imperfection $\epsilon$, indicating a phase transition from MBL phase to trivial thermal phase.

\begin{figure}[t]\centering
\includegraphics[width=0.4\textwidth]{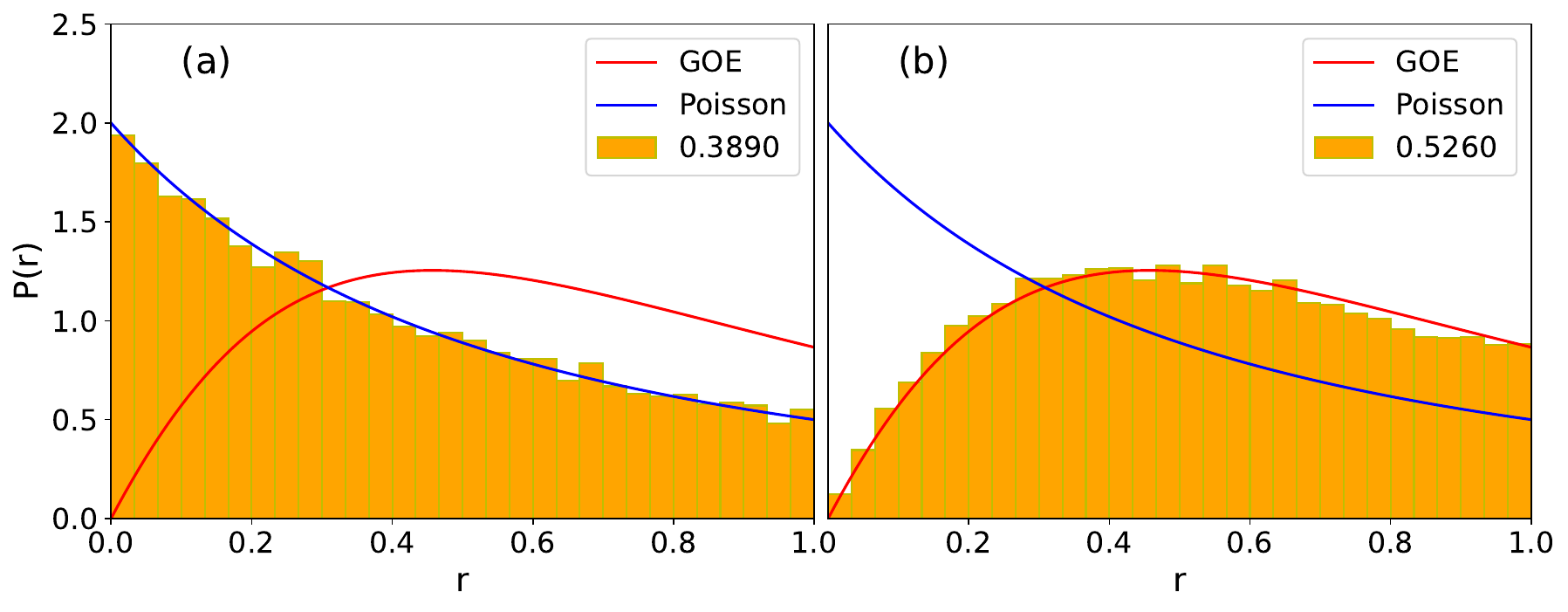}
\caption{Histogram of the distribution of the level spacing ratios: $N=15$, $J_{z}=\pi+1.5$, $W=5.0$. (a), $\epsilon= 0.05$; (b), $\epsilon=0.5$. With increasing the imperfection $\epsilon$, the distribution of level spacing ratio gradually crosses from the Poisson limit to the GOE distribution.}
\label{fig:his-r-N13-main}
\end{figure}

To diagnose the phase transition and discrete time translational symmetry breaking, we utilize two indicators: the magnitude of the subharmonic response and the mutual information between the first and the last site. In principle, the critical imperfection $\epsilon_{c}$ can be extracted from the data collapses for mutual information \cite{PhysRevLett.118.030401}. Due to limitation of the system size accessible, the critical value $\epsilon_{c}$ and the critical exponents can not be accurately determined for some Hamiltonian parameters (see more details on the finite-size data collapse in the SM \cite{Note1}). Nonetheless, the convincing numerical results indeed imply the existence of the DTC phase, see Fig. \ref{fig:diagram} for the schematic phase diagram. Even in the presence of a generic spin-spin interaction, the DTC phase still exists and is robust against the imperfection $\epsilon$ (see the SM for more numerical results \cite{Note1}).

\begin{figure}[t]\centering
	\includegraphics[width=0.35\textwidth]{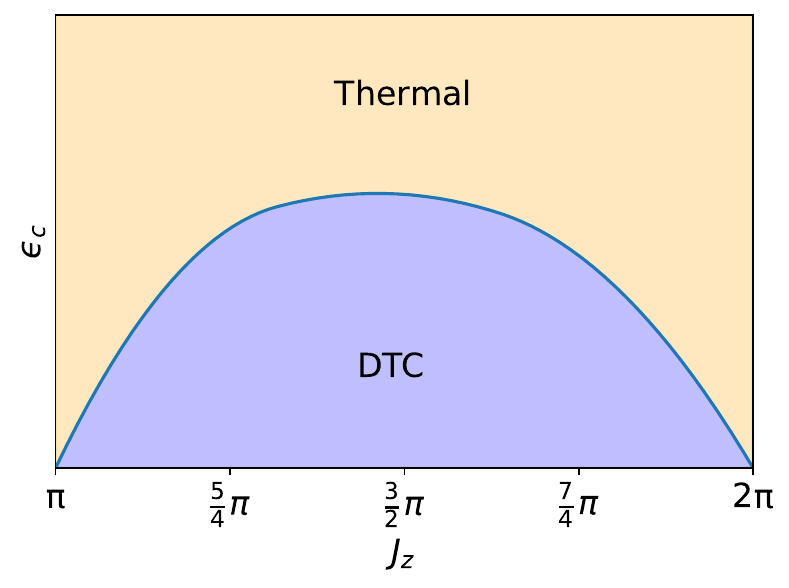}
	\caption{The schematic phase diagram: when $J_{z}=n\pi$, the $zz$ interaction has no effect on the Floquet evolution and the system enters trivial thermalized phase.}
	\label{fig:diagram}
\end{figure}

{\bf Discussions and Conclusion:}
It was reported that Stark many-body localization can be induced by strong external magnetic field even in the presence of a local phonon bath \cite{sarkar_protecting_2022}. On the contrary, DTC stabilized by random disorder MBL is unstable against environmental coupling in open systems \cite{DTC_open_system}. Due to the different mechanisms between random MBL and Stark MBL, an interesting future direction is to investigate whether the DTC phase enabled by Stark MBL is robust when coupling to the environment.

We have demonstrated that the discrete time crystal can be realized in a clean kicked Floquet model stabilized by Stark MBL. 
We also utilize the perturbation theory to explain the novel beating timescale absent in conventional DTC. Compared to the conventional DTC stabilized by the strong disorder, the resources required in our model are much fewer and it can be easily realized on the NISQ hardware \cite{Preskill2018quantumcomputingin, arute_quantum_2019} (see the SM for detailed experimental proposals \cite{Note1}).

~\newline
 \textbf{Acknowledgements:} We thank Zhou-Quan Wan for helpful discussions. This work is supported in part by the MOSTC Grants No. 2021YFA1400100 and No. 2018YFA0305604 (H. Y.), the NSFC under Grant No. 11825404 (S. X. Z., S. L., and H. Y.), the CAS Strategic Priority Research Program under Grant No. XDB28000000 (H. Y.), and Beijing Municipal Science and Technology Commission under Grant No. Z181100004218001 (H. Y.).



%

\clearpage

\begin{widetext}

\date{\today}
\maketitle


\section*{Supplemental Materials: Discrete time crystal enabled by Stark many-body localization}

	\renewcommand{\theequation}{S\arabic{equation}}
	\setcounter{equation}{0}
	\renewcommand{\thefigure}{S\arabic{figure}}
	\setcounter{figure}{0}

\tableofcontents
\titlecontents{section}
              [0.0cm]
              {}%
              {\contentslabel{3.5em}}%
              {}%
              {\titlerule*[0.5pc]{}\contentspage}%

\titlecontents{subsection}
              [0.1cm]
              {}%
              {\contentslabel{3.0em}}%
              {}%
              {\titlerule*[0.5pc]{}\contentspage}%

\titlecontents{subsubsection}
              [1.0cm]
              {}%
              {\contentslabel{2.5em}}%
              {}%
              {\titlerule*[0.5pc]{}\contentspage}%

	\subsection{A. Why is there a dominant-subleading $\pi$-pair pattern: A perturbation theory perspective}
	In this section, via degenerate perturbation theory, we explain the overlap pattern between the initial product states and the quasi-eigenstates of the Floquet operator $U_F$.
	
	The Floquet unitary $U_{F}$ reads:
	\begin{eqnarray}
	U_{F} &=& e^{-iH_{2}}e^{-i(\frac{\pi}{2}-\epsilon)H_{1}}= e^{-iH_{2}}e^{-i(\frac{\pi}{2}-\epsilon) \sum_{j}X_{j}} = e^{-iH_{2}}e^{-i\frac{\pi}{2} \sum_{j}X_{j}} e^{i \epsilon \sum_{j}X_{j}} = U_{F}^{0}e^{i \epsilon \sum_{j}X_{j}} ,  \\ \no
	\end{eqnarray}
	and by Taylor expansion on small imperfection $\epsilon$, we have:
	\begin{eqnarray}
    U_{F} &=& U_{F}^{0} \sum_{n} \frac{(i\epsilon)^{n}}{n!} (\sum_{j}X_{j})^{n} = U_{F}^{0} (I + i \epsilon \sum_{j}X_{j} + O(\epsilon^{2})) = U_{F}^{0} + i\epsilon U_{F}^{0} \sum_{j}X_{j} + O(\epsilon^{2}) \\ \no
      	  &\approx& U_{F}^{0} + U_{F}^{1}. \no
	\end{eqnarray}

    Assuming $\vert \phi \rangle$ and $\vert -\varphi \rangle$ are two quasi-eigenstates of $U^{0}_{F}$ with quasi-energy difference $\pi$, which is approximately the case for ``good initial states".
	\begin{eqnarray}
	U_{F}^{0} \vert \phi \rangle &=& e^{-i \varepsilon^{0}} \vert \phi \rangle \quad \rm{and} \quad  U_{F}^{0} \vert -\varphi \rangle = -e^{-i \varepsilon^{0}} \vert -\varphi \rangle,
	\end{eqnarray}
	then $U_{F}^{0} \vert \varphi \rangle = e^{-i \varepsilon^{0}} \vert \varphi \rangle$, where $\varepsilon^{0} = \frac{1}{2}(H_{2}(z_{\phi}) + H_{2}(-z_{\phi})) = \frac{1}{2}(H_{2}(z_{\varphi}) + H_{2}(-z_{\varphi})) + \pi \quad (\rm{mod} \, \, 2\pi)$, $z_{\phi}$ and $z_{\varphi}$ are the corresponding computational basis configurations from quasi-eigenstates. And
    \begin{eqnarray}
	\vert \phi \rangle &=& \frac{1}{\sqrt{2}} (e^{-i \frac{H_{2}(z_{\phi})}{2}} \vert z_{\phi} \rangle  + e^{-i \frac{H_{2}(-z_{\phi})}{2}} \vert -z_{\phi} \rangle), \\
	\vert \varphi \rangle &=& \frac{1}{\sqrt{2}} (e^{-i \frac{H_{2}(z_{\varphi})}{2}} \vert z_{\varphi} \rangle  - e^{-i \frac{H_{2}(-z_{\varphi})}{2}} \vert -z_{\varphi} \rangle),
	\end{eqnarray}
	$\vert z_{\phi} \rangle $ and $\vert z_{\varphi} \rangle$ are related by a Pauli-$X$ matrix on site $j$ (here we assume that the choice of $J_z$ makes the quasi-energy difference is closest to $\pi$ with flipping on $j$ site, in the main text we fix $j=\frac{N-1}{2}$ for simplicity),
	\begin{eqnarray}
	\vert z_\phi \rangle &=& X_{j} \vert z_\varphi \rangle,
	\end{eqnarray}
	and by convention we have,
	\begin{eqnarray}
	\vert z_\phi \rangle &=& \prod_{j} X_{j} \vert -z_\phi \rangle.
	\end{eqnarray}
   	The zero order perturbative quasi-eigenstate of $U_{F}$ reads
	\begin{eqnarray}
	\vert \psi^0 \rangle = a \vert \varphi \rangle + b \vert \phi \rangle.
	\end{eqnarray}
	Then
	\begin{eqnarray}
	(U_{F}^{0}-e^{-i \varepsilon^{0}}) \vert \psi^{1} \rangle &=& (E_{1} - U_{F}^{1}) \vert \psi^{0} \rangle \\ \no
	&=& (E_{1} - U_{F}^{1}) (a \vert \varphi \rangle + b \vert \phi \rangle),
	\end{eqnarray}
	where $\vert \psi^{1} \rangle$ is the first order perturbative quasi-eigenstate of $U_{F}$. It requires
	\begin{eqnarray}
	\label{eq:det}
	\rm{det} \begin{bmatrix}
	   -E_{1} & \langle \varphi \vert U_{F}^{1} \vert \phi \rangle  \\
	    \langle \phi \vert U_{F}^{1} \vert \varphi \rangle & -E_{1}
	  \end{bmatrix}
	=0.
	\end{eqnarray}
	According to
	\begin{eqnarray}
	\vert z_{\phi} \rangle &=& \frac{1}{\sqrt{2}} e^{i \frac{H_{2}(z_{\phi})}{2}} (\vert \phi \rangle + \vert -\phi \rangle), \\
	\vert -z_{\phi} \rangle &=& \frac{1}{\sqrt{2}} e^{i \frac{H_{2}(-z_{\phi})}{2}} (\vert \phi \rangle - \vert -\phi \rangle),
	\end{eqnarray}

	\begin{eqnarray}
	\vert z_{\varphi} \rangle &=& \frac{1}{\sqrt{2}} e^{i \frac{H_{2}(z_{\varphi})}{2}} (\vert \varphi \rangle + \vert -\varphi \rangle), \\
	\vert -z_{\varphi} \rangle &=& \frac{1}{\sqrt{2}} e^{i \frac{H_{2}(-z_{\varphi})}{2}} (\vert -\varphi \rangle - \vert \varphi \rangle),
	\end{eqnarray}
	the off-diagonal elements can be calculated:
	\begin{eqnarray}
	\label{eq:off12}
	\langle \varphi \vert U_{F}^{1} \vert \phi \rangle &=& \langle \varphi \vert i\epsilon U_{F}^{0} \vert \frac{1}{\sqrt{2}}(e^{-i \frac{H_{2}(z_{\phi})}{2}} \vert z_{\varphi} \rangle  + e^{-i \frac{H_{2}(-z_{\phi})}{2}} \vert -z_{\varphi} \rangle) \\  \no
	&=&\langle \varphi \vert i\epsilon U_{F}^{0} \vert (\frac{1}{2}e^{-i \frac{H_{2}(z_{\phi})}{2}} e^{i \frac{H_{2}(z_{\varphi})}{2}} (\vert \varphi \rangle + \vert -\varphi \rangle) + \frac{1}{2} e^{-i \frac{H_{2}(-z_{\phi})}{2}} e^{i \frac{H_{2}(-z_{\varphi})}{2}} (\vert -\varphi \rangle - \vert \varphi \rangle) ) \\ \no
	&=& \frac{i\epsilon}{2}(e^{-i\frac{H_{2}(z_{\phi})}{2}}e^{i \frac{H_{2}(z_{\varphi})}{2}} e^{-i\varepsilon^{0}} -  e^{-i\frac{H_{2}(-z_{\phi})}{2}}e^{i \frac{H_{2}(-z_{\varphi})}{2}} e^{-i\varepsilon^{0}}) \\ \no
	&=& \frac{i\epsilon}{2}(-e^{-i\frac{H_{2}(z_{\phi})}{2}}e^{-i \frac{H_{2}(-z_{\varphi})}{2}} +  e^{-i\frac{H_{2}(-z_{\phi})}{2}}e^{-i \frac{H_{2}(z_{\varphi})}{2}} ),
	\end{eqnarray}

	 \begin{eqnarray}
	 \label{eq:off21}
	\langle \phi \vert U_{F}^{1} \vert \varphi \rangle &=& \langle \phi \vert i\epsilon U_{F}^{0} \vert \frac{1}{\sqrt{2}}(e^{-i \frac{H_{2}(z_{\varphi})}{2}} \vert z_{\phi} \rangle  - e^{-i \frac{H_{2}(-z_{\varphi})}{2}} \vert -z_{\phi} \rangle) \\ \no
	&=&\langle \phi \vert i\epsilon U_{F}^{0} \vert (\frac{1}{2}e^{-i \frac{H_{2}(z_{\varphi})}{2}} e^{i \frac{H_{2}(z_{\phi})}{2}} (\vert \phi \rangle + \vert -\phi \rangle) - \frac{1}{2} e^{-i \frac{H_{2}(-z_{\varphi})}{2}} e^{i \frac{H_{2}(-z_{\phi})}{2}} (\vert \phi \rangle - \vert -\phi \rangle) ) \\ \no
	&=& \frac{i\epsilon}{2}(e^{-i\frac{H_{2}(z_{\varphi})}{2}}e^{i \frac{H_{2}(z_{\phi})}{2}} e^{-i\varepsilon^{0}} -  e^{-i\frac{H_{2}(-z_{\varphi})}{2}}e^{i \frac{H_{2}(-z_{\phi})}{2}} e^{-i\varepsilon^{0}}) \\ \no
	&=& \frac{i\epsilon}{2}(e^{-i\frac{H_{2}(z_{\varphi})}{2}}e^{-i \frac{H_{2}(-z_{\phi})}{2}}  -  e^{-i\frac{H_{2}(-z_{\varphi})}{2}}e^{-i \frac{H_{2}(z_{\phi})}{2}} ).
	\end{eqnarray}

	Bring the results of Eq. \ref{eq:off12} and Eq. \ref{eq:off21} back to Eq. \ref{eq:det}, we have
	\begin{eqnarray}
	E_{1}^{2} + \frac{\epsilon^{2}}{4} e^{-2i\varepsilon^{0}} (2- e^{-i (H_{2z}(z_\phi) + H_{2z}(-z_\varphi))} - e^{-i (H_{2z}(-z_\phi) + H_{2z}(z_\varphi))} ) &=& 0 , \\
	E_{1}^{2} + \frac{\epsilon^{2}}{4} e^{-2i\varepsilon^{0}} (2- e^{-2ijW} - e^{+2ijW} ) &=& 0 , \\
	E_{1}^{2} + \frac{1}{2}\epsilon^{2} e^{-2i\varepsilon^{0}} (1-\cos(2jW) ) &=& 0 ,\\
	E_{1}^{2} + \epsilon^{2} e^{-2i\varepsilon^{0}} \sin^{2}(jW)  &=& 0 ,
	\end{eqnarray}
	where $H_{2z}$ is the linear Zeeman field term. And the first order perturbation correction to quasi-eigenenergy is
	\begin{eqnarray}
	E_{1} = \pm i \epsilon e^{-i\varepsilon^{0}} \sin(jW),
	\end{eqnarray}
	where $j$ is the site position for the matching flip $X_j$ and $W$ is the linear Zeeman field strength. Then the quasi-eigenenergy are
	\begin{eqnarray}
	e^{-i \varepsilon^{\prime}} &=& e^{-i\varepsilon^{0}}(1 \pm i \epsilon \sin(jW)),  \\ \no
	&\approx& e^{-i\varepsilon^{0}} e^{\pm i \epsilon \sin(jW)},
	\end{eqnarray}
	\begin{eqnarray}
	\varepsilon^{\prime} = \varepsilon^{0} \pm  \epsilon \sin(jW)  \quad (\rm{mod} \, \, 2\pi).
	\end{eqnarray}
	So $\omega_{b} = 2 \epsilon \sin(jW)$
	, i.e. the second timescale $T_{b} = \frac{2\pi}{\omega_{b}}= \frac{\pi}{\epsilon \sin(jW)}$ . And $T_{b}$ is the same for all ``good initial states''  since the derivation above doesn't depend on special good state configurations (see Fig. \ref{fig:deltaef}). This result leads a dominant beating oscillation in the dynamics of the state-averaged fidelity. The predictions from perturbation theory and numerical results are in good agreement, see Fig. \ref{fig:per}.
	
 \begin{figure}[H]\centering
	\includegraphics[width=0.40\textwidth]{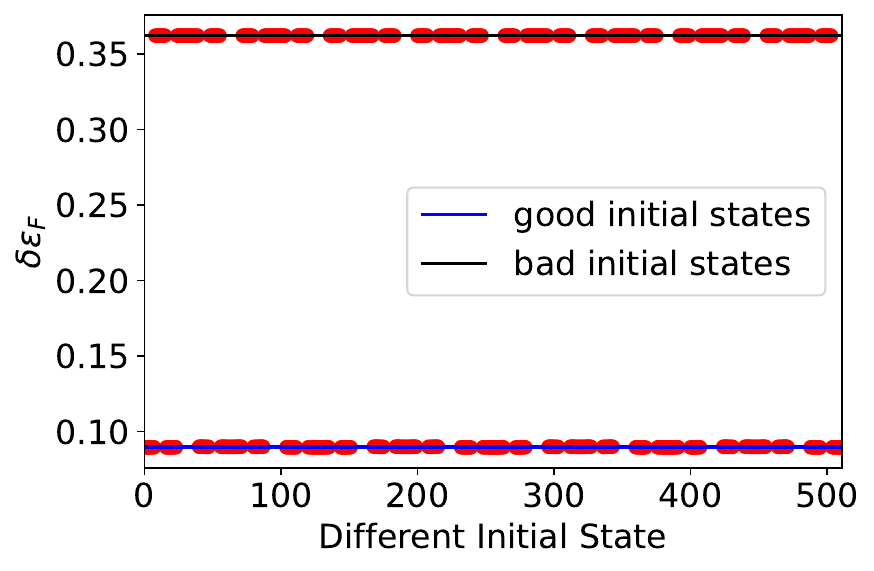}
	\caption{The quasi-eigenenergy differences between dominant and subleading $\pi$-pair $\delta \varepsilon_{F}$ ($T_{b}$) are the same for all ``good initial states''. For ``bad initial states'', although there is no obvious subleading $\pi$-pair, we can also calculate the $\delta \varepsilon_{F}$ according to the magnitude of the overlaps and it is roughly determined by $2J_{z}$. It is consistent with our understanding from perturbation picture. ($N=9, J_{z}=\frac{\pi}{2N}, W=5.0, \epsilon=0.05$)}
	\label{fig:deltaef}
\end{figure}

	\begin{figure}[H]
	\centering
	\begin{minipage}[c]{0.5\textwidth}
	\centering
	\includegraphics[height=5.0cm,width=7.5cm]{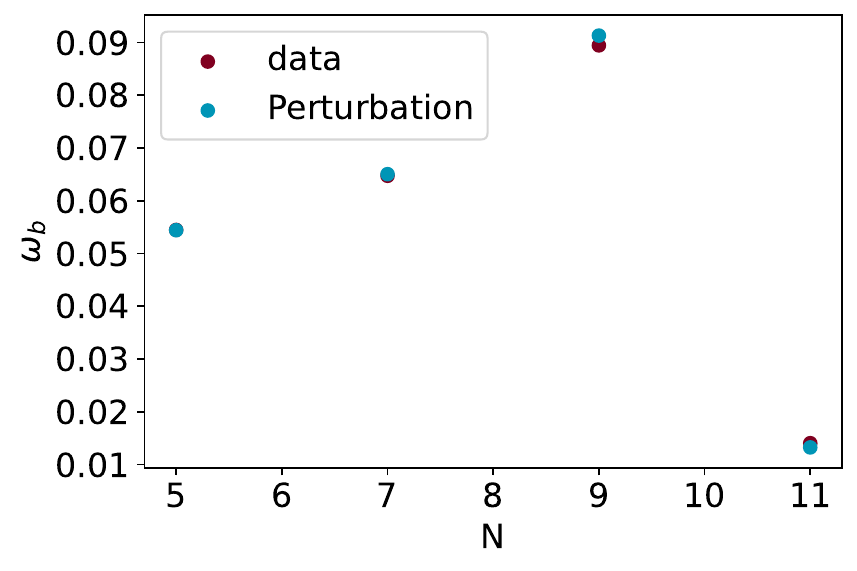}
	\end{minipage}%
	\begin{minipage}[c]{0.5\textwidth}
	\centering
	\includegraphics[height=5.0cm,width=7.5cm]{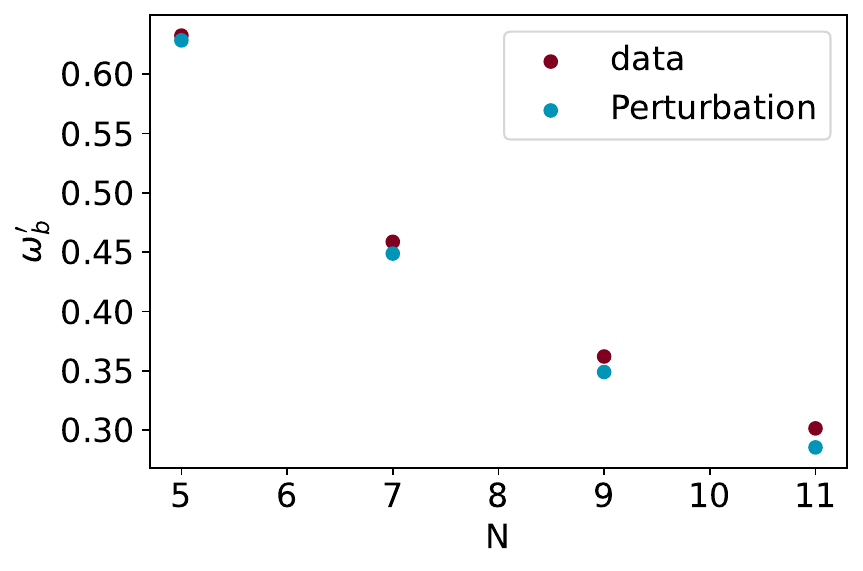}
	\end{minipage}
	\caption{The beating oscillation frequency is determined by Fourier transformation on even period of dynamics using state-averaged fidelity mentioned in the main text. As discussed in the text, there are two different frequencies, $\omega_{b}$ and $\omega_{b}^{\prime}$. For ``good initial states'', $\omega_{b} \approx \frac{2\pi}{T_{b}}=2 \epsilon \sin(\frac{N-1}{2}W)$; for ``bad initial states'', $\omega_{b}^{\prime} \approx 2J_{z}$. Left: $\omega_{b}$; Right: $\omega_{b}^{\prime}$. ($J_{z}=\frac{\pi}{2N}; W=5.0; \epsilon=0.05$)}.
	\label{fig:per}
	\end{figure}

	When the quasienergy difference between two quasi-eigenstates of $U_{F}^{0}$ equals zero instead of $\pi$. Then the first order perturbative quasi-eigenenergy are:
	\begin{eqnarray}
	E_{1}^{2} + \frac{1}{2}\epsilon^{2} e^{-2i\varepsilon^{0}} (1+\cos(2jW) ) &=& 0 ,\\
	E_{1}^{2} + \epsilon^{2} e^{-2i\varepsilon^{0}} \cos^{2}(jW)  &=& 0 ,
	\end{eqnarray}
	\begin{eqnarray}
	\varepsilon^{\prime} = \varepsilon^{0} \pm   \epsilon \cos(jW)  \quad (\rm{mod} \, \, 2\pi).
	\end{eqnarray}

	When we choose a $J_{z}$ that is not perfect, i.e. $\delta_{J_{z}} > 0$ ($J_{z}=\frac{\pi-\delta_{J_{z}}}{2N}$), there are no quasi-eigenstates of $U_{F}^{0}$ with the quasi-eigenenergy difference exactly equals $0$ or $\pi$ besides intrinsic $\pi$-pairs. For each $\pi$-pair up to one X flip on the target one, the quasi-eigenenergy difference is $\pi-\delta_{J_{z}}$, and the one with the smallest $\delta_{J_{z}}$ constitutes the subleading $\pi$-pair in the overlap spectrum. Therefore, we can still see a dominant beating timescale $T_{b}$ determined by the quasi-eigenenergy difference between dominant and subleading $\pi$-pairs.

	For ``bad initial states'', we show why there is another beating timescale $T_{b}^{\prime}$ different from the $T_{b}$ for ``good initial state". For a given bad initial state (assuming matching site $j=2$ here), for example, $\vert z \rangle = \vert 01100 \rangle $, $\vert z \rangle $ and $\vert -z \rangle$ form a $\pi$-pair of quasi-eigenstates of $U_{F}^{0}$, if we flip spin $S_{j}$ $(j=0,...N)$ of $\vert z \rangle$ and $\vert -z \rangle$, there will be $N$ new $\pi$-pairs of quasi-eigenstates of $U_{F}^{0}$. In this case, different $\pi$-pairs are still related by the Pauli-X matrix and the second timescale $T_{b}^{\prime}$ is roughly determined by the original quasi-eigenenergy difference. For example, we set $J_{z} = \frac{\pi}{2N}$, when $N$ is large, $J_{z} \sim 0$, related quasi-eigenstates with quasi-eigenenergy difference of $2J_{z}$ correspond to the $T_{b}^{\prime}$, see Fig. \ref{fig:per}. The difference is that there are more than one subleading $\pi$-pairs for a given dominant $\pi$-pair. Also from the perturbation theory, we can see many different beating frequencies in such cases.

	In the above, we choose $J_{z}$ depending on the system size $N$ to obtain the beating timescale analytically for simplicity. For a more general size independent $J_{z}$, the novel beating timescale may become less evident. Nonetheless, the DTC phase still exists and is robust against the imperfection $\epsilon$.

	\subsection{B. Analysis of different timescales}
	\subsubsection{The relationship between the second timescale and quasi-energy difference}

	A random initial product state $\vert z \rangle$ can be roughly written as a superposition of a dominant and a subleading $\pi$ pairs,
	\begin{eqnarray}
	\vert z \rangle = \alpha (\vert m \rangle + \vert -m \rangle) + \beta (\vert n \rangle + \vert -n \rangle ),
	\end{eqnarray}
	where $\vert\alpha \vert^2+\vert\beta\vert^2= \frac{1}{2} $, and $U_{F} \vert \pm m \rangle = \pm e^{- i\varepsilon_{m}} \vert m \rangle$, $U_{F} \vert \pm n \rangle = \pm e^{- i\varepsilon_{n}} \vert n \rangle$. At time $t=2lT$.
	\begin{eqnarray}
	U_{F}^{2l} \vert z \rangle = e^{-2li(\varepsilon_{m}-\varepsilon_{n})} \alpha (\vert m \rangle + \vert -m \rangle) + \beta (\vert n \rangle + \vert -n \rangle),
	\end{eqnarray}
	and
	\begin{eqnarray}
	F_{2l} = \vert \langle z \vert U_{F}^{2l} \vert z \rangle \vert^{2} = 4\vert \alpha \vert^{4}+4\vert \beta \vert^{4}+8\vert \alpha \vert^{2} \vert \beta \vert^{2} \cos(2l(\varepsilon_{m}-\varepsilon_{n})).
	\end{eqnarray}
	Then the second timescale $T_{b}$, or the frequence is related the quasi-energy difference:
	\begin{eqnarray}
	\omega_{b} = \frac{2\pi}{T_{b}} = \frac{2\pi}{2l} = \vert \varepsilon_{m}-\varepsilon_{n} \vert.
	\end{eqnarray}

    \subsubsection{The relationship between the third timescale and the quasienergy mismatch}

	For a finite-size system, there is a tiny quasi-energy mismatch in each $\pi$-pair and this is the origin of the third timescale.
	\begin{eqnarray}
	\vert z \rangle = \frac{1}{\sqrt{2}}(\vert m \rangle + \vert -(m+\delta_{m}) \rangle),
	\end{eqnarray}
	At time $t=2lT$,
	\begin{eqnarray}
	U_{F}^{2l} \vert z \rangle = \frac{1}{\sqrt{2}}(\vert m \rangle + e^{-2li\delta_{m}} \vert -(m+\delta_{m}) \rangle),
	\end{eqnarray}
	then the third timescale for the given initial state $\vert z \rangle$ is:
	\begin{eqnarray}
	T_{g}= 2l = \frac{2\pi}{\delta_{m}}.
	\end{eqnarray}

	\subsubsection{The dominant beating timescale in the dynamics of state-averaged fidelity}
	In Fig. \ref{fig:state6-and-average} for state-averaged fidelity, we can see a dominant beating timescale $T_{b}$, which is caused by the ``good initial states'', and many other beating timescales $T_{b}^{\prime}$, which are caused by ``bad initial states''. We explain why the $T_{b}$ is dominant in state averaged case. Consider the fidelity of a given initial state firstly:
	\begin{eqnarray}
	F_{n} = \vert \langle \psi \vert U_{F}^{n} \vert \psi \rangle \vert^{2},
	\end{eqnarray}
	where $\vert \psi \rangle$ is the initial state. And the initial state can be roughly represented by $n$ $\pi$-pairs
	\begin{eqnarray}
	\vert \psi \rangle = \sum_{i=0}^{n-1} \alpha_{i}(\vert + \rangle_{i} + \vert - \rangle_{i}),
	\end{eqnarray}
	where $2\sum_{i=0}^{n-1} \vert \alpha_{i} \vert^{2}=1$. At time $t=2lT$,
	\begin{eqnarray}
	U_{F}^{2l} \vert \psi \rangle = \sum_{i=0}^{n-1} \alpha_{i} e^{-2li \delta \varepsilon_{F}(i,0)} (\vert + \rangle_{i} + \vert - \rangle_{i}),
	\end{eqnarray}
	where $\delta \varepsilon_{F}(i,0)$ is the quasi-eigenenergy difference between $\vert + \rangle_{i}$ and $\vert + \rangle_{0}$. And
	\begin{eqnarray}
	F_{2l} &=& \vert \sum_{i=0}^{n-1} 2 \vert \alpha_{i} \vert^{2} e^{-2li \delta \varepsilon_{F}(i,0)} \vert^{2} \\
    &=& \sum_{i=0}^{n-1} 4 \vert \alpha_{i} \vert^{4} + \sum_{i < j}^{n-1} 8 \vert \alpha_{i} \vert^{2} \vert \alpha_{j} \vert^{2} \cos(2l(\delta \varepsilon_{F}(i,0) - \delta \varepsilon_{F}(j,0))) .
	\end{eqnarray}
	The total height of the peak after Fourier transformation is:
	\begin{eqnarray}
	\rm{Height} \propto \sum_{i < j}^{n-1} 8 \vert \alpha_{i} \vert^{2} \vert \alpha_{j} \vert^{2} = 1 - \sum_{i=0}^{n-1}4 \vert \alpha_{i} \vert^{4} = 1 - 4\rm{EIPR},
	\end{eqnarray}
	where $\rm{EIPR}= \sum_{i} \vert \alpha_{i} \vert^{4}$ \cite{santagati_witnessing_2018, liu_probing_2021}. For ``bad initial state'', there is a dominant $\pi$-pair and $\rm{EIPR}$ is large; for ``good initial state'', there is a dominant $\pi$-pair and a subleading $\pi$-pair and $\rm{EIPR}$ is relatively small. Since the numbers of the ``good initial states'' and ``bad initial states'' are both the same: $2^{N-1}$, the dynamics of the state-averaged fidelity will exhibit a dominant beating timescale as given by the ``good initial states''.

	\subsection{C. The Fourier transform on the fidelity dynamics}
	To see the oscillation period in the dynamics more obviously, we can take Fourier transform of fidelity, which is defined as:
	\begin{eqnarray}
	f(\omega)= \sum_{l=0}^{n-1}\bar{F}_{l} e^{-i \omega l},
	\end{eqnarray}
	where $\bar{F}_{l}$ is the state-averaged fidelity in the time step $l$. And we take the normalized form of $f(\omega)$ in this Letter, which reads:
	\begin{eqnarray}
	\tilde{f}(\omega) = \frac{\vert f(\omega) \vert}{n}.
	\end{eqnarray}
	For example, the height of the peak of subharmonic response is $\tilde{f}(\pi)$.
	The spectrum can show some symmetries as we elaborated below. Since the dynamics of fidelity at odd periods are zero and the dynamics of fidelity at even periods can be regarded as a sine function for simplicity, the fidelity at $n$-th period is:
	\begin{eqnarray}
	\label{fidelity}
	F_{n} = A(1+\sin(\pi n + \frac{\pi}{2})) + B(\sin(\omega_b n) - \sin((\pi-\omega_b)n) + \sin((\pi + \omega_b)n) - \sin((2\pi-\omega_b)n)),
	\end{eqnarray}
	where the first term corresponds to the subharmonic response and the second term corresponds to the beating oscillation. Therefore, we can see four peaks in the spectrum with frequencies $\omega_b$, $\pi - \omega_b$, $\pi + \omega_b$, and $2\pi - \omega_b$, corresponding to the beating oscillation, which are consistent with the results shown in  Fig. \textcolor{red}{3} and Fig. \textcolor{red}{4} in the main text.


	\subsection{D. The numerical results for different imperfection $\epsilon$}
	To distinguish different phases controlled by the imperfection $\epsilon$, we can utilize the magnitude, $\tilde{f}(\pi)$, of the peak of the subharmonic response. When $\epsilon$ is small, there is an obvious subharmonic response for DTTSB. With increasing the imperfection $\epsilon$, the magnitude of the subharmonic response decreases. When $\epsilon$ is large enough, the fidelity at even periods will decay to zero after a few dynamical cycles and there is no obvious signal of the subharmonic response. The numerical results are summarized in Fig. \ref{fig:diffe}.

	\begin{figure}[H]\centering
		\includegraphics[width=0.8\textwidth]{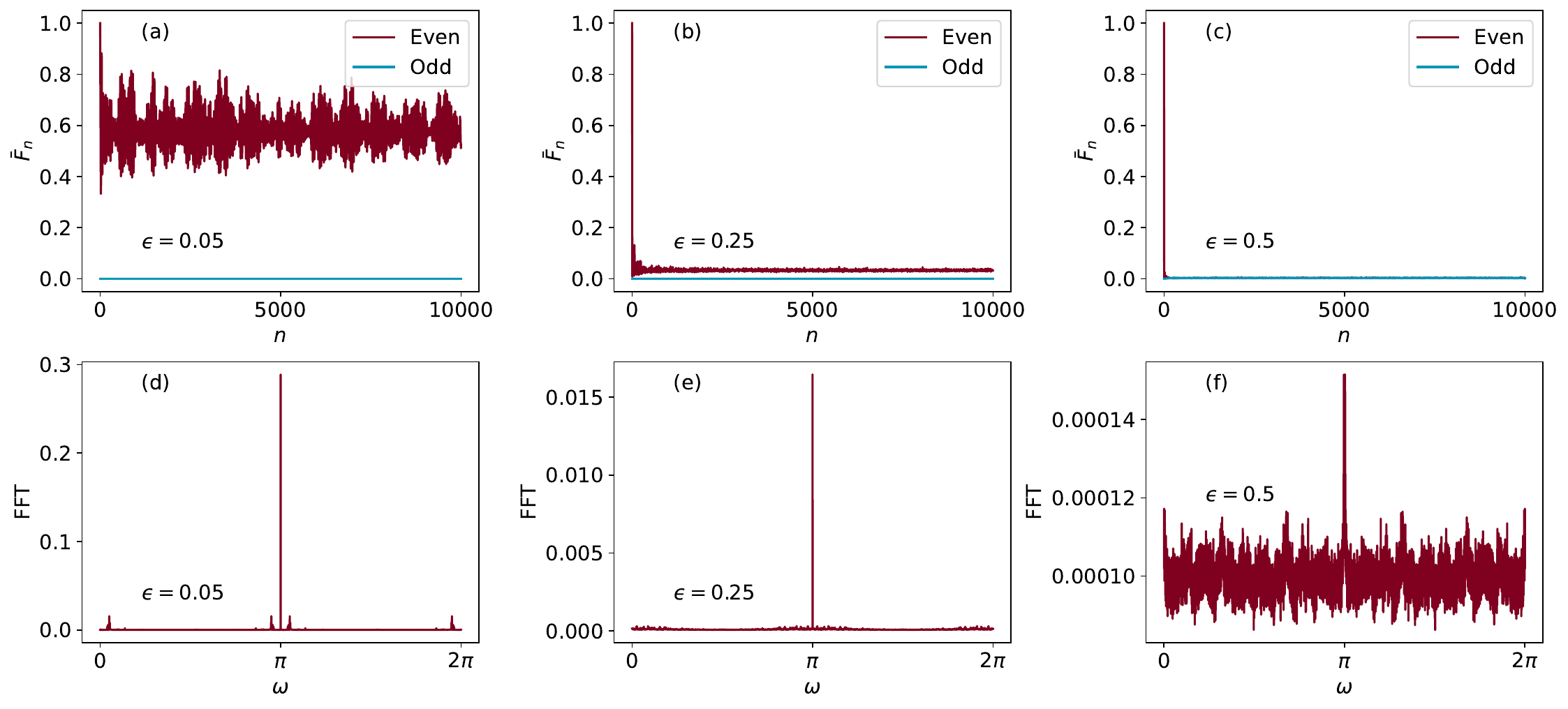}
		\caption{State-averaged fidelity and its Fourier transformation $\bar{F}_{n}$: (a)(d), $\epsilon=0.05$; (b)(e), $\epsilon=0.25$; (c)(f), $\epsilon=0.5$. Here, $N=11$, $J_{z}=\pi+1.5$, $W=5.0$.}
		\label{fig:diffe}
	\end{figure}

	\subsection{E. Level statistics for many-body localization}
	Level statistics are often used to diagnose MBL phases.
	The level spacing ratio for quasi-eigenenergies is defined as:
	\begin{eqnarray}
	r_{n}=\frac{\rm{min}\{\Delta^{n},\Delta^{n+1}\}}{\rm{max}\{\Delta^{n},\Delta^{n+1}\}},
	\end{eqnarray}
	where $\Delta^{n} = \varepsilon^{n}-\varepsilon^{n+1}$ is the gap between quasi-eigenenergy levels $n$ and $n+1$, $[r^{(n)}]$ is the level average of this ratio. In the delocalized phase (thermal phase), the level spacings follow a GOE distribution and the level-averaged level spacing ratio is $[r^{(n)}] \approx 0.53$ \cite{PhysRevLett.110.084101}. In contrast, in the localized phase, the level spacing follows a Poisson distribution, which gives $[r^{(n)}] = 2 \ln2 - 1 \approx 0.39$ \cite{PhysRevLett.110.084101}.

	As discussed in the main text, the DTC order is stabilized by Stark MBL. We can see the distribution of level spacing ratio gradually crosses from the Poisson limit to the GOE type with increasing imperfection $\epsilon$, see Fig. \ref{fig:levelspacingratio} and Fig. \ref{fig:his-r-N13} for level spacing distribution.

	\begin{figure}[H]\centering
	\includegraphics[width=0.40\textwidth]{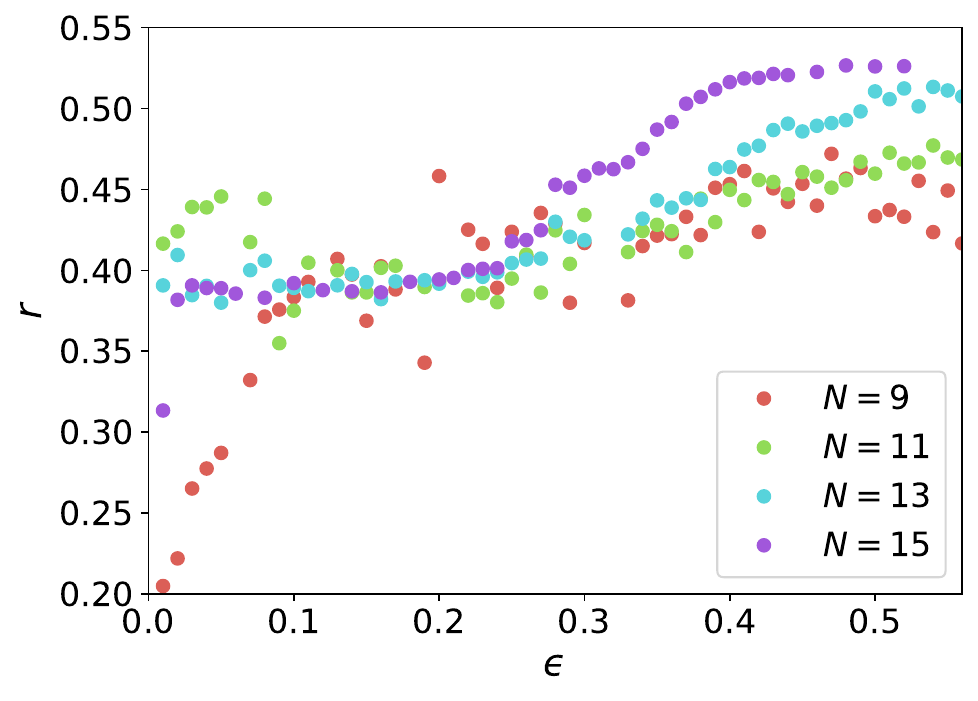}
	\caption{Level spacing ratio: $J_{z}=\pi+1.5$, $W=5.0$.}
	\label{fig:levelspacingratio}
	\end{figure}


	\begin{figure}[t]\centering
	\includegraphics[width=0.45\textwidth]{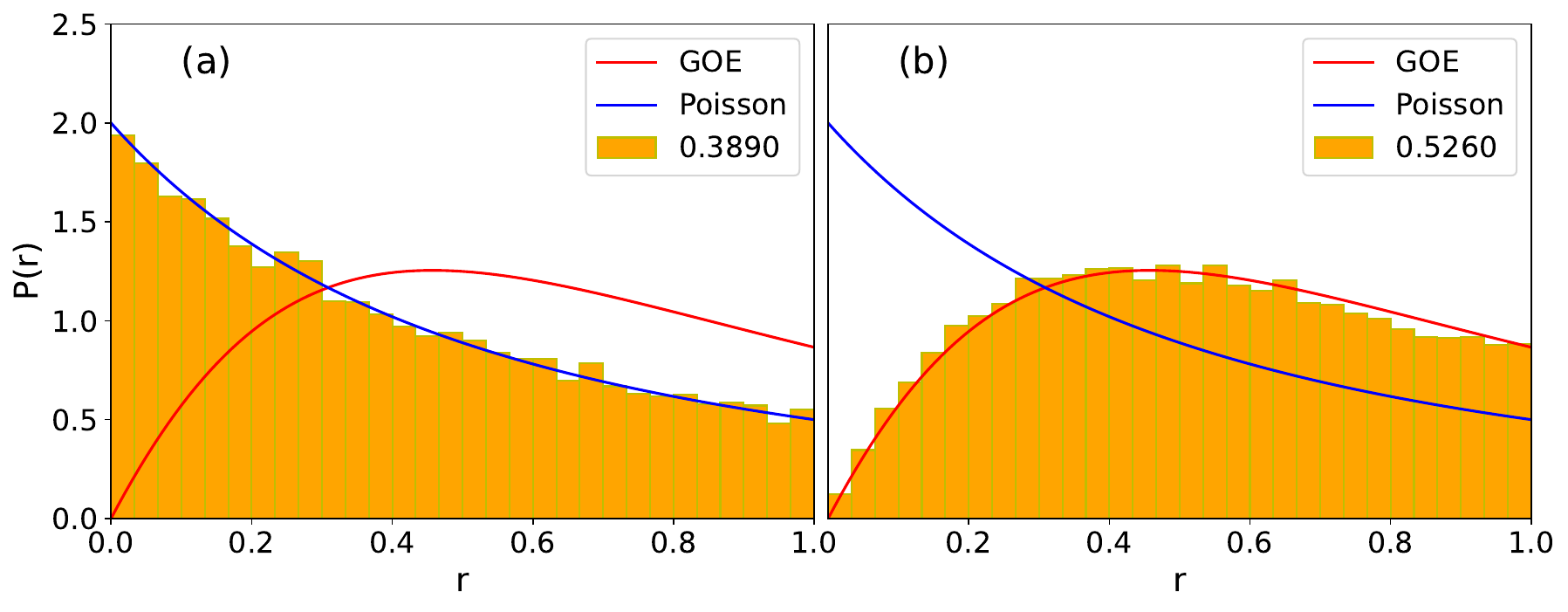}
	\caption{Histogram of the distribution of the level spacing ratios: $N=15$, $J_{z}=\pi+1.5$, $W=5.0$. (a), $\epsilon= 0.05$; (b), $\epsilon=0.5$. With increasing the imperfection $\epsilon$, the distribution of level spacing ratio gradually crosses from the Poisson limit to the GOE distribution.}
	\label{fig:his-r-N13}
	\end{figure}

	\subsection{F. Different choices of $J_{z}$ and $W$}
	As discussed in \cite{mi_time-crystalline_2021}, the linear Zeeman field, as well as the Stark MBL, will be suppressed by the imperfection $\epsilon$ and a special term of $zz$ interaction is important (linear form in our case) to protect the MBL phase. In this section, we show the results of different combinations of $J_{z}$ and $W$. We find that DTC indeed requires the linear $zz$ interaction, otherwise the impact of linear Zeeman field is suppressed. There are four choices of $\{J_{z}$, $W \}$: (a), $\{$constant, constant$\}$; (b), $\{$constant, linear$\}$; (c), $\{$linear, constant$\}$; (d), $\{$linear, linear$\}$. For example, $\{$constant, linear$\}$ stands for
	\begin{eqnarray}
	H_{2}=J_{z}\sum_{j}Z_{j}Z_{j+1} + W\sum_{j}jZ_{j}.
	\end{eqnarray}
	The results of the Fourier peak height $\tilde{f}(\pi)$ with different choices $\{J_{z}$, W$\}$ are shown in Fig. \ref{fig:different-choices}. And we choose an arbitrary $J_z$ with no size dependence to demonstrate the universality of the DTC phase and the physical picture in this Letter.

	\begin{figure}[H]
	\centering
	\begin{minipage}[c]{0.5\textwidth}
	\centering
	\includegraphics[height=4.0cm,width=6.5cm]{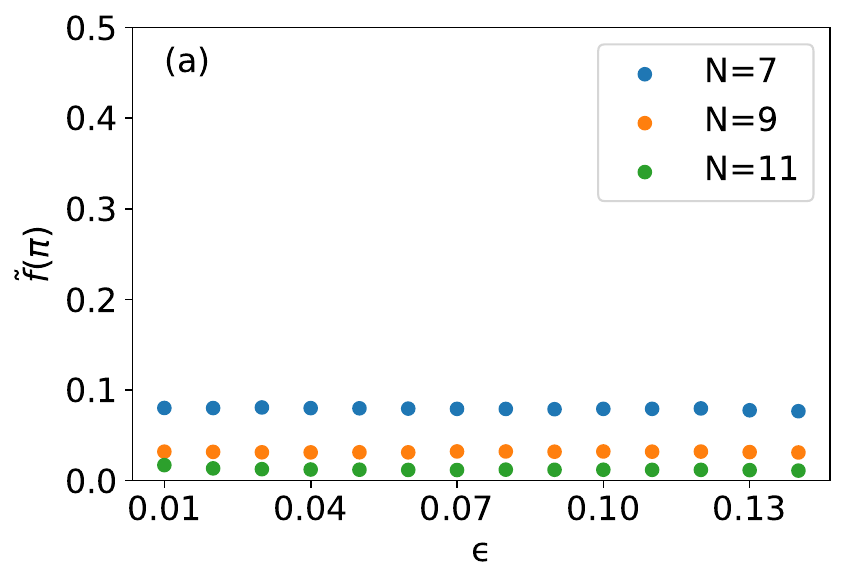}
	\end{minipage}%
	\begin{minipage}[c]{0.4\textwidth}
	\centering
	\includegraphics[height=4.0cm,width=6.5cm]{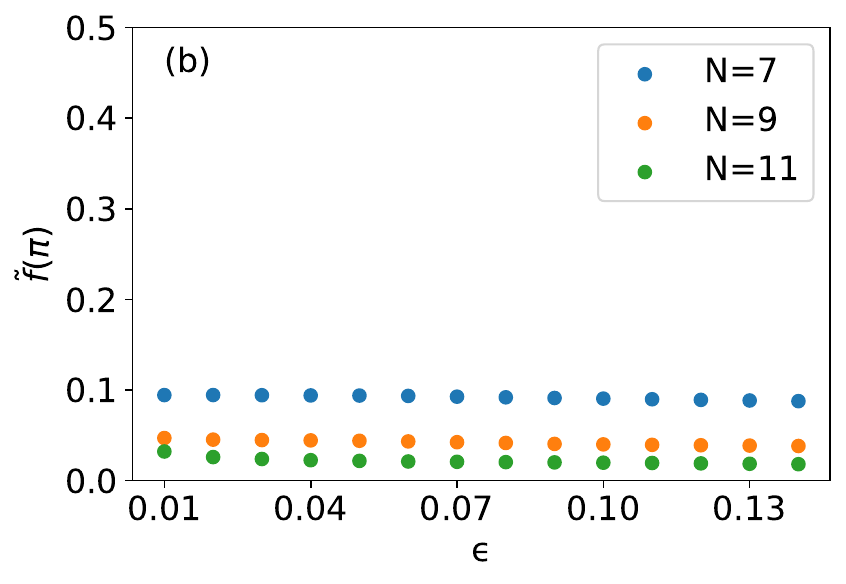}
	\end{minipage}
	\begin{minipage}[c]{0.5\textwidth}
	\centering
	\includegraphics[height=4.0cm,width=6.5cm]{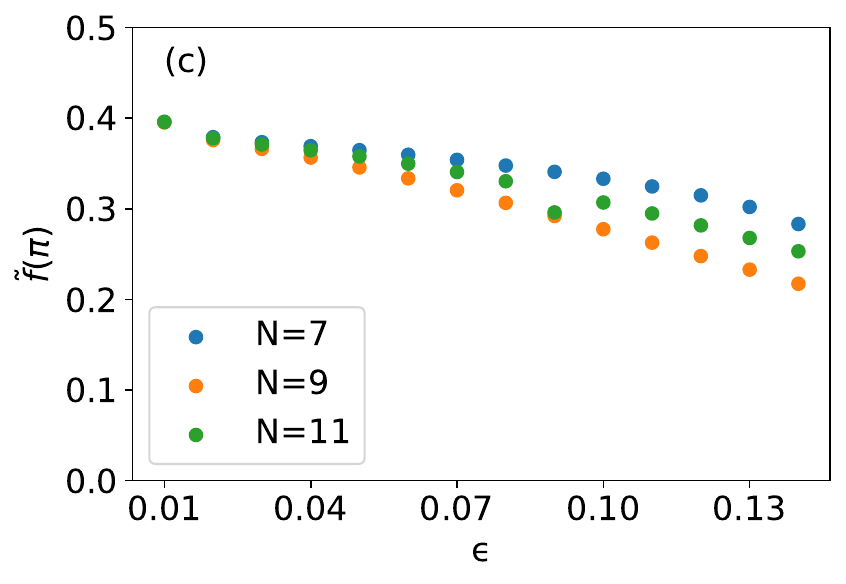}
	\end{minipage}%
	\begin{minipage}[c]{0.4\textwidth}
	\centering
	\includegraphics[height=4.0cm,width=6.5cm]{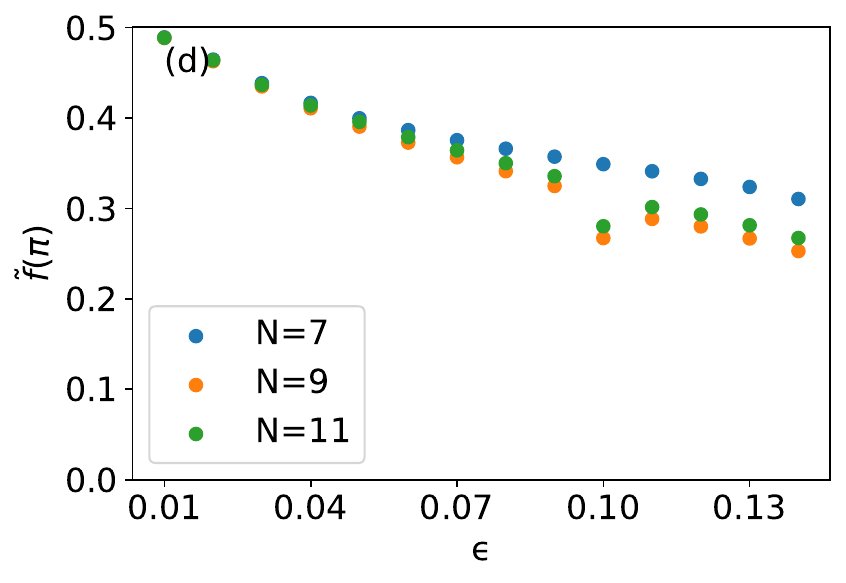}
	\end{minipage}%
	\caption{The height $\tilde{f}(\pi)$ of the subharmonic response with different choices of $J_{z}$ and $W$: (a), $\{$constant, constant$\}$; (b), $\{$constant, linear$\}$; (c), $\{$linear, constant$\}$; (d), $\{$linear, linear$\}$. With the constant $zz$ interaction, we find that the subharmonic response is not robust against the imperfection $\epsilon$ as shown in (a) and (b). And the linear $zz$ interaction is important to stabilize the DTC as shown in (c) and (d). The difference between constant $W$ and linear $W$,  whose impact is suppressed by the imperfection $\epsilon$, is not obvious ($J_{z}=\pi+1$; $W=5.0$).}
	\label{fig:different-choices}
	\end{figure}

	\subsection{G. The dynamics of autocorrelator}
	Besides fidelity discussed in the main text, we can also utilize autocorrelator as the observable to study the Floquet dynamics which is easier to obtain from experiments. All conclusions obtained in the main text, including DTC response and beating timescale, still hold for autocorrelators.
	The autocorrelator is defined as:
	\begin{eqnarray}
	\overline{\langle Z(0)Z(n) \rangle} = \frac{1}{N}\sum_{j} \langle Z_j(0)Z_j(nT) \rangle.
	\end{eqnarray}
	The autocorrelator dynamics starting from a ``good initial state'' are shown in Fig. \ref{fig:gsautocorr}. The autocorrelator dynamics starting from a ``bad initial state'' and the state-averaged autocorrelator dynamics are shown in Fig. \ref{fig:bsautocorr}.

	Different from the fidelity (see Eq. \ref{fidelity}), the autocorrelator at $n$-th period is in the form (considering only the dominant frequencies):
	\begin{eqnarray}
	\label{corr}
	F_{n} = && A\sin(\pi n + \frac{\pi}{2}) \\ \nonumber
	&+& B(\sin(\omega_b n) - \sin((\pi-\omega_b)n) + \sin((\pi + \omega_b)n) - \sin((2\pi-\omega_b)n)) \\ \nonumber
	&-& B(\sin(\omega_b n) + \sin((\pi-\omega_b)n) - \sin((\pi + \omega_b)n) - \sin((2\pi-\omega_b)n)),
	\end{eqnarray}
	where the first term corresponds to the subharmonic response, the second term corresponds to the beating oscillation at even periods, and the third term corresponds to the beating oscillation at odd periods. Therefore,
	\begin{eqnarray}
	\label{corr}
	F_{n} = && A\sin(\pi n + \frac{\pi}{2}) + 2B( - \sin((\pi-\omega_b)n) + \sin((\pi + \omega_b)n) ).
	\end{eqnarray}
	We can see two peaks corresponding to the beating oscillation with frequencies $\pi-\omega_b$ and $\pi+\omega_b$, as shown in Fig. \ref{fig:gsautocorr} and Fig. \ref{fig:bsautocorr}. To facilitate the analysis of the beating timescale, we can calculate the frequency spectrum of autocorrelators at only even periods as shown in the insets of Fig. \ref{fig:gsautocorr} and Fig. \ref{fig:bsautocorr}. Although the signal of the beating oscillation is much weaker than that of the DTC response, the beating timescale still exists and is in the same value as that obtained from the fidelity dynamics (see Fig. \textcolor{red}{3} and Fig. \textcolor{red}{4}).

	\begin{figure}[H]\centering
		\includegraphics[width=1.0\textwidth]{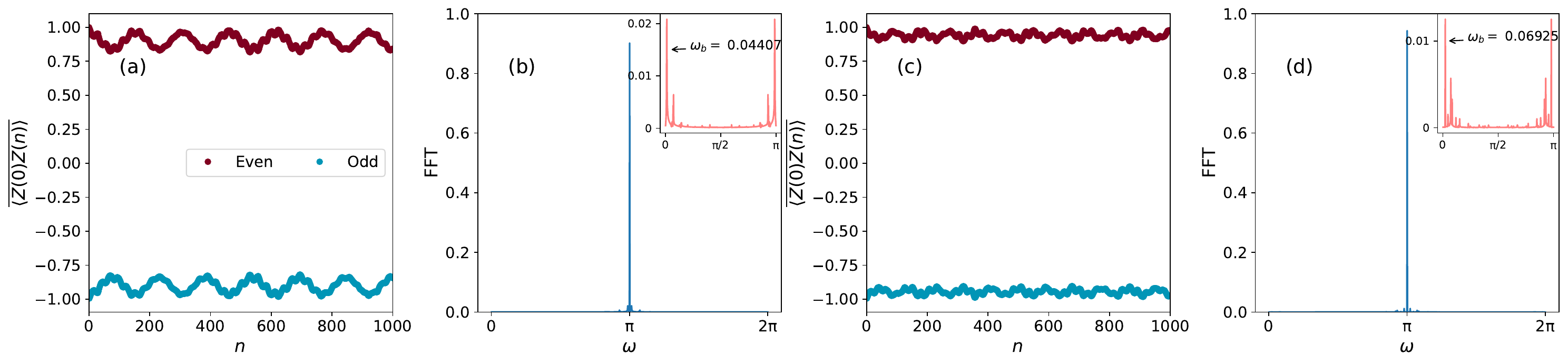}
		\caption{Autocorrelator dynamics of the product state $\vert 000000000000000 \rangle$ which is a ``good initial state'' ($N=15$; $W=5.0$; $\epsilon=0.05$). In (a)(b), $J_{z} = \frac{\pi}{2N}$; in (c)(d), $J_{z}= \frac{\pi-0.05}{2N}$. The insets are the frequency spectrums of the autocorrelators at even periods. The beating timescales are the same as those obtained from the fidelity dynamics (see Fig. \textcolor{red}{3}).}
		\label{fig:gsautocorr}
	\end{figure}

	\begin{figure}[H]\centering
		\includegraphics[width=1.0\textwidth]{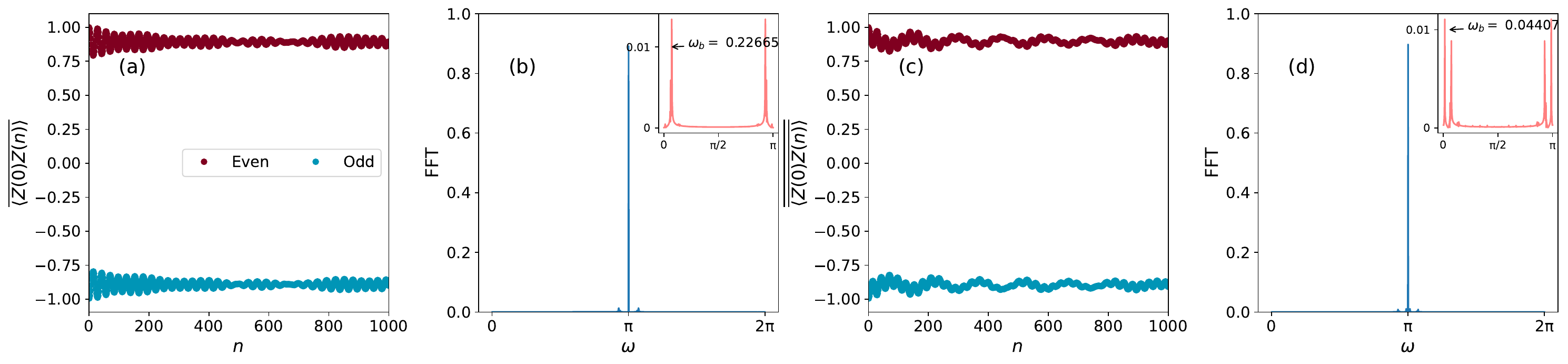}
		\caption{(a)(b): Autocorrelator dynamics of the product state $\vert 001100110011001 \rangle$ ($N=15; J_{z}=\frac{\pi}{2N}; W=5.0; \epsilon=0.05$). It is a bad initial state and there is another dominant beating timescale. (c)(d): dynamics of the state-averaged autocorrelator ($N=15; J_{z}=\frac{\pi}{2N}; W=5.0; \epsilon=0.05$). There is a dominant beating timescale $T_{b}=\frac{2\pi}{\omega_{b}}$ which is induced by the ``good initial states''. We can also see many other timescales caused by the ``bad initial states''.}
		\label{fig:bsautocorr}
	\end{figure}

	\subsection{H. Phase transition and the impact of the generic interaction}

	In the main text, we focus on the clean kicked Floquet model with linear $zz$ interaction (see Eq. \ref{zzH2}). In this section, we investigate the dynamics and phase transition of a generalized Floquet model in the presence of a generic spin-spin interaction,
	\bea
	H_{2} = J \sum_{j}(X_{j}X_{j+1} + Y_{j}Y_{j+1}) + J_{z} \sum_{j}(j+1)Z_{j}Z_{j+1} + W\sum_{j} j Z_{j}.
	\eea
	$J=0$ corresponds to the model studied in the main text. We use this generalized model to demonstrate the universal properties of the new DTC model.

    To diagnose the phase transition and discrete time translational symmetry breaking, we utilize two indicators: the magnitude of the subharmonic response and the mutual information. The former is related to the definition of $DTTSB$-1 in \cite{PhysRevLett.117.090402}: DTTSB occurs when the observables develop persistent oscillations whose periods are an integer multiple of the driving period; the latter is related to the definition of $DTTSB$-2 in \cite{PhysRevLett.117.090402}: DTTSB occurs if the eigenstates of the Floquet unitary $U_{F}$ cannot be short-range correlated. Due to that the DTC phase is stabilized by MBL, we also compute the level spacing ratio. The numerical results from different $J_z$ are summarized in Fig. \ref{fig:XY}.

	As we increase the imperfection $\epsilon$, the magnitude, $\tilde{f}(\pi)$, of the peak of subharmonic response decreases, see Fig. \ref{fig:XY}(c)(f), and eventually, becomes completely washed out when the system enters into the trivial thermal phase.

	\begin{figure}[H]\centering
		\includegraphics[width=0.8\textwidth]{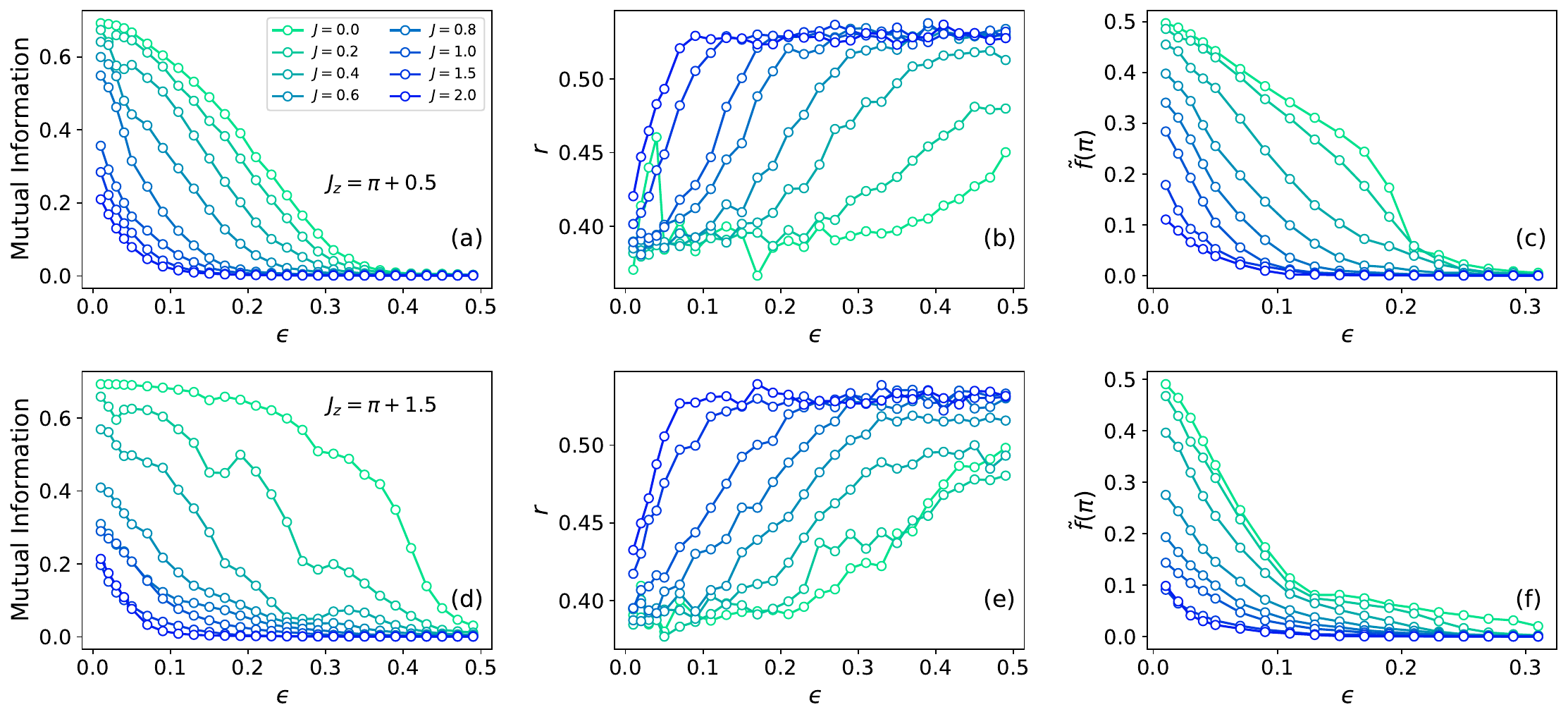}
		\caption{(a), Mutual information ($J_{z}=\pi+0.5$, $N=13$); (b), level spacing ratio $r$ ($J_{z}=\pi+0.5$, $N=13$); (c), Fourier spectrum peak height $\tilde{f} (\pi)$ ($J_{z}=\pi+0.5$, $N=9$); (d), Mutual information ($J_{z}=\pi+1.5$, $N=13$); (e), level spacing ratio $r$ ($J_{z}=\pi+1.5$, $N=13$); (f), Fourier spectrum peak height $\tilde{f} (\pi)$ ($J_{z}=\pi+1.5$, $N=9$).}
		\label{fig:XY}
	\end{figure}

	We further use the state-averaged mutual information $I(A, B)=S_{A}+S_{B}-S_{AB}$ \cite{PhysRevLett.100.070502, jian_long-range_2015, PhysRevLett.117.090402, PhysRevLett.118.030401}, to check whether the quasi-eigenstate is short-range entangled. Here $A$ and $B$ are spins on opposite ends of a $N$-spin chain and $S$ is the Von Neumann Entropy. For small imperfection $\epsilon \approx 0$, any quasi-eigenstate is long-range entangled ``cat state'', indicating nearly full $I(A, B)= \rm{log}2$, and mutual information drops dramatically upon leaving the DTTSB phase for large $\epsilon$, see Fig. \ref{fig:XY}(a)(d).
	
	With increasing imperfection $\epsilon$, the distribution of level spacing ratio also gradually crosses from the Poisson limit to the GOE type, see Fig. \ref{fig:XY}(b)(e).

	To further verify the existence of the DTC phase, we can determine the critical imperfection $\epsilon_c$ and corresponding critical exponents via the data collapse with the scaling function proposed in Ref. \cite{PhysRevLett.118.030401},	
	\bea
	I(N, \epsilon) = N^{-\beta} F((\epsilon-\epsilon_c)N^{1/\nu}),
	\eea
	where $\beta$ and $\nu$ are the critical exponents. See Fig. \ref{fig:collapse} for the data collapse of the numerical results in the presence of the generic interaction ($J=0.8$). The critical imperfection $\epsilon_c=0.04(0.07)$ for $J_{z}=\pi+0.5(\pi+1.5)$ respectively and the critical exponents predicted are consistent with those reported in Ref. \cite{PhysRevLett.118.030401}.

	All these convincing results demonstrate the existence of the DTC phase. With increasing the strength of the generic spin-spin interaction $J$, the mutual information and the magnitude of the subharmonic response $ \tilde{f}(\pi)$ both decrease, and the level spacing ratio $r$ gets closer to the value predicted by the GOE distribution. These facts indicate that the critical imperfection $\epsilon_c$ may be suppressed by the generic interaction $J$. However, when the generic interaction is relatively weak, the critical imperfection $\epsilon_{c}>0$, i.e.,  the DTC phase still exists and is robust against the imperfection as indicated in Fig. \ref{fig:collapse}.

	Note that due to the limitation of the system size accessible, the data collapse is affected by the finite-size effect and thus the phase diagram in the main text is only schematic.



	\begin{figure}[H]\centering
		\includegraphics[width=0.65\textwidth]{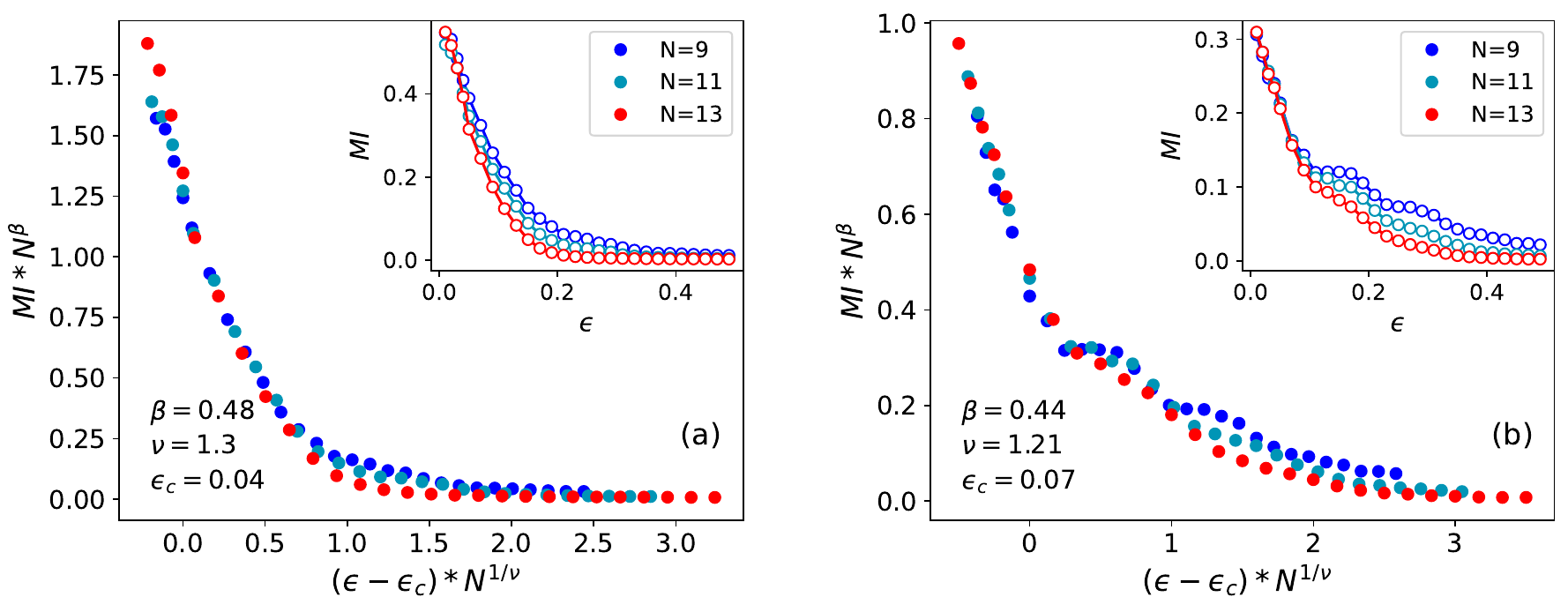}
		\caption{Data collapse: $J_{z}=\pi+0.5$ (left) and $J_{z}=\pi+1.5$ (right). Here, $J=0.8$, $W=5.0$. In the presence of the relatively weak generic interaction, the critical imperfection $\epsilon_c$ is nonzero and the DTC phase still exists below $\epsilon_c$. And the critical exponents are consistent with those in Ref. \cite{PhysRevLett.118.030401}.}
		\label{fig:collapse}
	\end{figure}

	\subsection{I. The realization of DTC on quantum computers}
	Our model can be easily realized on the NISQ hardware with significantly fewer resources and less requirements on SPAM error. In this section, we will show the experimental realization and protocol of our model on quantum devices. Our proposed quantum circuit structure for the DTC consists of three parts: the preparation circuit, Floquet unitary evolution circuit, and the measurement part, as shown in Fig. \ref{fig:circuit}. $R_{a}(\theta)$ is the rotation gate along $a$ axis ($a=x,y,z$). For example, $R_{x}(\theta)=\rm{exp}(-iX\theta/2)$, where $X$ is the Pauli matrix. And $R_{zz}(\theta)$ is parameterized $zz$ coupling gate defined as $R_{zz}(\theta)=\rm{exp}(-i\frac{\theta}{2} Z \otimes Z)$. For a given product state $\vert z \rangle$, we can get many computational basis configurations after measuring the output state of the quantum circuit on computational basis and the probability of getting back the original $\vert z \rangle$ configuration is the fidelity we focus on. Besides, we can also measure site averaged spin polarization as the dynamics indicator.

	\begin{figure}[H]\centering
	\includegraphics[width=0.65\textwidth]{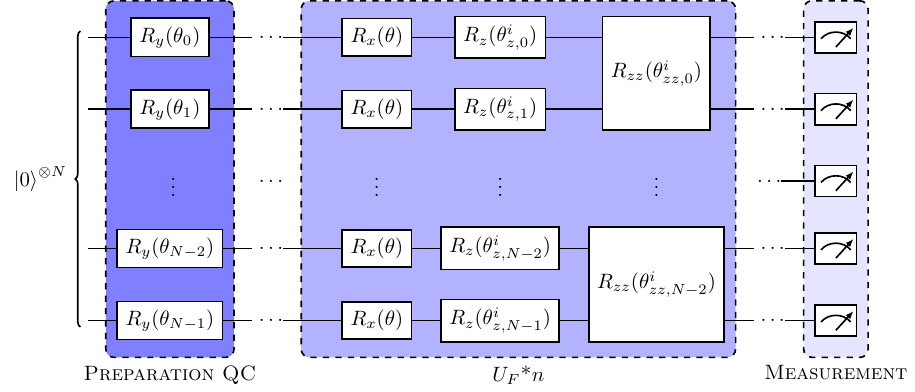}
	\caption{The circuit structure for the implementation of kicked model in this Letter. In the preparation circuit part, we first prepare a product state $\vert z \rangle = \vert S_{0}, S_{1},...S_{N-1} \rangle$ from the initial state $\vert 000...\rangle$ with $\theta_{j}= S_{j}\pi$ ($S=0$: spin up; $S=1$: spin down). In the Floquet unitary evolution part, $R_{x}$ corresponds to the kicked term $U_{1}$; $R_{z}$ is the term of linear Zeeman field and $R_{zz}$ is the term of linear $zz$ interaction. In the measurement part, we will get a computational basis configuration $\vert z^{\prime} \rangle$. The probability of getting $\vert z \rangle$ is the fidelity ($\theta=\pi-2\epsilon$, $\theta^{i}_{z,j}=2jW$, $\theta^{i}_{zz,j}=2(j+1)J_{z}$).}
	\label{fig:circuit}
	\end{figure}

\end{widetext}

\end{document}